\documentclass{article}
\usepackage[english]{babel}
\usepackage{amsmath,amssymb,tabularx,bbm,bm,booktabs}
\usepackage{xcolor,slashed,epsfig,cite}
\usepackage{verbatim,slashed,ulem,geometry,setspace,algorithm}
\usepackage{algorithmic,mathrsfs,braket,multirow,dcolumn}
\usepackage{graphicx}
\graphicspath{{./Figures/}}
\usepackage[colorlinks=true,citecolor=blue,urlcolor=cyan,filecolor=magenta,linkcolor=red,frenchlinks=true,bookmarks]{hyperref}
\allowdisplaybreaks[4]
\newcommand{\nobracket}{}

\newcommand{\tmop}[1]{\ensuremath{\operatorname{#1}}}

\renewcommand\sout{\bgroup \color{red}  \ULdepth=-.5ex \ULset}
\definecolor{darkyellow}{rgb}{0.55,1,0.87}

\begin{document}

\title{Generalized parton distributions of $\Delta$ resonance in a diquark spectator approach}

\author{Dongyan Fu$^{a,b,\thanks{fudongyan@ihep.ac.cn}}$, 
Bao-Dong Sun$^{c,d,e, \thanks{bao-dong.sun@m.scnu.edu.cn}}$,
and Yubing Dong$^{a,b,\thanks{dongyb@ihep.ac.cn}}$\\ 
Institute of High Energy Physics, Chinese Academy 
of Sciences, Beijing 100049, China$^{a}$\\
School of Physical Sciences, University of Chinese 
Academy of Sciences, Beijing 101408, China$^{b}$\\
Guangdong Provincial Key Laboratory of Nuclear Science,
Institute of Quantum Matter,\\
South China Normal University,  Guangzhou 510006, China$^{c}$\\
Guangdong-Hong Kong Joint Laboratory of Quantum Matter,\\
Southern Nuclear Science Computing Center, \\
South China Normal University, Guangzhou 510006, China$^{d}$\\
Institut f\"ur Theoretische Physik II, Ruhr-Universit\"at Bochum, D-44780 Bochum, Germany$^{e}$
}

\newenvironment{mysubeq}
{\begin{subequations}
\renewcommand\theequation{\theparentequation 
\alph{equation}}}
{\end{subequations}}
\vspace{1em}
\maketitle
\begin{abstract}
The generalized parton distributions (GPDs) for the spin-3/2 $\Delta^+$ resonance are studied numerically by using a diquark spectator approach. Our results show that symmetric constraints from time reversal on GPDs are satisfied. The axial vector 
form factors of the system are also provided and compared with the lattice QCD calculation. Furthermore, the structure functions are obtained from GPDs in the forward limit. The evolution of structure functions to the scales up to 4 GeV are carried out as predictions for the possible lattice QCD calculations. 

\end{abstract}

\section{Introduction}\label{Introduction}

\quad The usual parton distribution functions (PDFs), like $f(x)$, describe the 
longitudinal momentum distribution of partons inside a hadron. The generalized 
parton distributions (GPDs), $H_i(x,\xi,t)$, with two additional variables: the 
skewness $\xi$ and momentum transfer $t$, embody much richer nonperturbative 
information of the partonic structures inside a hadron, and thus provide more 
comprehensive descriptions of hadron structures in terms of partons. 
The Mellin moments (i.e. $x$-moments) of GPDs can produce the electromagnetic form 
factors (EMFFs) and gravitational form factors (GFFs), i.e. the form factors of the electromagnetic vector and the energy momentum tensor.
The latter provides important information on the fundamental 
mechanical properties of the hadron like mass, spin, and internal force 
distributions (pressure and shear forces). Since the concept of GPDs was 
introduced\cite{Muller:1994ses}, many efforts have been devoted to nucleon 
GPDs~\cite{Polyakov:2002yz} which are accessible by semi-inclusive processes 
through deeply virtual Compton scattering (DVCS), and deeply virtual meson 
production (DVMP) \cite{Ji:1996ek,Ji:1996nm,Radyushkin:1997ki}. It  
attracts increasing attention in recent years with the DVCS having been 
measured by the CLAS collaboration at JLab. \cite{CLAS:2018ddh,CLAS:2021gwi} 
and the COMPASS collaboration at CERN \cite{Schnell:2012zz}, and there are planned experiments 
at the future electron ion colliders (EIC, EIcC) 
\cite{AbdulKhalek:2021gbh,Anderle:2021wcy}. It should be mentioned that the EIC 
or EIcC may also have spin 3/2 candidate targets like $^{7}_{3}$Li, $^{9}_{4}$Be, 
and $^{11}_{4}$B.   
The factorization of GPDs for spin-3/2 particles has been given in our previous work~\cite{Fu:2022bpf}. 
The lowest mass spin-3/2 particles are $\Delta$ resonances and we will take $\Delta^+$ as an example to introduce. In contrast to the electromagnetic properties of the $\Delta^+$ resonance which have been extensively studied both experimentally and  theoretically\cite{ParticleDataGroup:2022pth,Alexandrou:2008bn,Fu:2022rkn}, it's difficult to extract the  GPDs of $\Delta$ isobars experimentally because of the short-lived nature of $\Delta$ resonances. Therefore, it is expected that the lattice QCD calculations are more practical to obtain model-independent estimations.

As far as we know, there are no estimates have been made yet for the GPDs of spin-3/2 particles, either from model calculations or the lattice QCD approach. 
The MIT lattice group has calculated the gluonic GFFs for $\Delta$ resonances \cite{Pefkou:2021fni} that can be related to GPDs through the sum rules. 
Although it's usually more expansive to calculate the form factors of quark than gluonic ones on the lattice, it still seems more promising than the experiment to get the model-independent results for the quark sector. It's interesting to employ model calculations to estimate how these GPDs behave concerning their three variables. In this work, we make such a numerical calculation using the diquark spectator approach. The dependence on all three variables of GPDs is shown explicitly and the symmetric properties with respect to skewness $\xi$ are verified. In the quark model picture, the gluon contribution is considered as being absorbed into the constituent quarks with the larger mass, and the model is considered reasonable at a rather lower renormalization scale than that of the usual lattice QCD calculations or experiments. Note that GPDs are scale-dependent quantities and the gluonic part would appear as the scale increases. The evolution is needed to compare the quark model results with the experimental data or lattice QCD calculations. The QCDNUM package~\cite{Botje:2010ay} is adopted to perform the next-to-next-to-leading order (NNLO) DGLAP evolution equation to make predictions for the structure functions up to a few GeV scales.\footnote{In preparation of the paper, there appeared one paper \cite{Lappi:2023lmi} that discussed the direct evolution of structure functions instead of through its relation with PDFs, where the former doesn't need to define the factorization scheme as the latter. Its toy-model result shows that the only visible difference between the two approaches is in the small $x$ region which is beyond the consideration of this work.} The evolution of the $x$-moments of the $d$ quark unpolarized structure function $F_1^d(x)$ are also calculated which may be accessible on the lattice.

\quad This work is organized as follows:
In Sec.~\ref{defination}, we briefly restate the expressions for the definitions of GPDs of the spin-3/2 system and the relations with EMFFs and GFFs which are given in our previous work \cite{Fu:2022bpf}. 
In Sec.~\ref{model}, the diquark spectator approach used in this work is specified. The corresponding numerical results, including GPDs, axial vector form factors, structure functions, and the related evolution, are shown in Sec.~\ref{result}.
Sec.~\ref{summary} is devoted to a short summary and discussion.\\

\section{Generalized parton distributions of a spin-3/2 particle}\label{defination}
Here we briefly show the conventions for defining GPDs of the spin-3/2 particles, 
which has been given in detail in our previous work~\cite{Fu:2022bpf}. 
The notations for the momenta are
\begin{eqnarray} 
P = \frac{p + p'}{2}, \quad q = p' - p, \quad t = q^2, 
\end{eqnarray}
where $p$ and $p'$ are the initial and final momenta, respectively.
The conventions of variables, skewness $\xi$ and $x$, in the representations of GPDs are
\begin{equation}
	\xi = - \frac{q \cdot n}{2 P \cdot n} = - 
	\frac{q^+}{2P^+} \quad  (| \xi | \leqslant 1), 
	\quad \text{and} \quad x = \frac{k \cdot n}{P \cdot n} 
	= \frac{k^+}{P^+}
	\quad  (-1 \leqslant x \leqslant 1),
\end{equation}
where $n$ is the light-like four vector $n = (n^+, n^-, \bm{n}_\perp) = (0, 2, \mathbf{0})$ with $n^2=0$. The scalar product of two four-vectors is $u \cdot v = \frac{1}{2} u^+ v^- + \frac{1}{2} u^- v^+ - \mathbf{u} \cdot \mathbf{v}$.\\

In  Ref.~\cite{Fu:2022bpf}, the quark GPDs of the  spin-3/2 particle, $H_i(x,\xi,t)$ and ${\tilde H}_i(x,\xi,t)$, are given as
\begin{equation}
	\begin{split}
	  V_{\lambda' \lambda}  =& \frac{1}{2} 
	  \int \frac{\text{d} z^-}{2 \pi} e^{i x (P \cdot z)}
	  \left.\left\langle p', \lambda'\left| 
	  \overline{\psi} \left( - z/2
	  \right) \slashed{n} \psi \left(  z/2
	  \right)\right| p, \lambda \right\rangle 
	  \right|_{z^+ = 0, \mathbf{z} = \mathbf{0}}\\
	 =& -\overline{u}_{\alpha'} (p', \lambda')
	   \mathcal{H}^{\alpha' \alpha}
	  (x, \xi, t) u_{\alpha} (p, \lambda) \\
     =& -\overline{u}_{\alpha'} (p', \lambda') \Bigg\{
    H_1 \frac{g^{\alpha' \alpha}}{M} + H_2 
     \frac{P^{\alpha'} P^{\alpha}}{M^3} + H_3  
     \frac{n^{[\nobracket \alpha'}
     P^{\alpha \nobracket]}}{M P \cdot n} + H_4  \left[\frac{3 M n^{\alpha'}
     n^{\alpha}}{(P \cdot n)^2} + \frac{g^{\alpha' \alpha}}{M}\right] + H_5  \left[\frac{g^{\alpha'
     \alpha} \slashed{n}}{P \cdot n}-\frac{g^{\alpha' \alpha}}{M}\right]\\
     & + H_6  \frac{P^{\alpha'} P^{\alpha}\slashed{n}}{M^2 P
     \cdot n} + H_7  \frac{n^{[\alpha' \nobracket}
     P^{\nobracket \alpha]} \slashed{n}}{(P \cdot n)^2} 
     + H_8 \left[ \frac{3 M^2 n^{\alpha'} n^{\alpha} 
     \slashed{n}}{(P \cdot n)^3} - \frac{3 M n^{\alpha'}
     n^{\alpha}}{(P \cdot n)^2} \right]
    \Bigg\} u_{\alpha} (p, \lambda)
   \end{split} \label{unGPDDefine}
\end{equation}
for unpolarized case, and 
\begin{equation}
	\begin{split}
	  A_{\lambda' \lambda}  = & \frac{1}{2} 
	  \int \frac{\text{d} z^-}{2 \pi}e^{i x (P \cdot z)}
	  \left. \left\langle p', \lambda' \left| \overline{\psi} 
	  \left( -  z/2 \right) \slashed{n} \gamma_5 
	  \psi \left(  z/2 \right) \right| p,
	  \lambda \right\rangle \right|_{z^+ = 0, \mathbf{z} = \mathbf{0}}\\
	=& - \overline{u}_{\alpha'}(p', \lambda')
	  \mathcal{\tilde H}^{\alpha' \alpha}
	  (x, \xi, t) u_{\alpha} (p, \lambda) \\
    =& - \overline{u}_{\alpha'}(p', \lambda') \Bigg\{
    \tilde{H}_1 \frac{g^{\alpha' \alpha}}{M} \gamma^5 + \tilde{H}_2 
    \frac{P^{\alpha'} P^{\alpha}}{M^3} \gamma^5 
    + \tilde{H}_3  \frac{n^{\left\{ \alpha \right.'}
    P^{\left.\alpha \right\}}}{M P \cdot n}\gamma^5 
    + \tilde{H}_4  \frac{M n^{\alpha'}
    n^{\alpha}}{(P \cdot n)^2} \gamma^5 + \tilde{H}_5  
    \frac{3 g^{\alpha' \alpha}}{\sqrt{5} P \cdot n} \slashed{n} \gamma^5\\
    & + \tilde{H}_6  \frac{3 P^{\alpha'} P^{\alpha}}{\sqrt{5} M^2 \left(P
    \cdot n \right)} \slashed{n} \gamma^5 + \tilde{H}_7  
    \frac{ n^{\left\{ \alpha \right.'}
    P^{\left.\alpha \right\}}}{(P \cdot n)^2}\slashed{n} 
    \gamma^5 + \tilde{H}_8  \left[ \frac{
    \sqrt{5} M^2 n^{\alpha'} n^{\alpha}}{(P \cdot n)^3} +\frac{ g^{\alpha' \alpha}}{\sqrt{5} P \cdot n} \right] \slashed{n} \gamma^5 \Bigg\} u_{\alpha} (p, \lambda),
	\end{split} \label{poGPDDefine}
\end{equation}
for polarized case, where $H_i \equiv H_i(x,\xi,t)$ and $\tilde{H}_i \equiv \tilde{H}_i(x,\xi,t)$
and $\lambda (\lambda')$ is the helicity of the initial (final) state. 
The normalization of the Rarita-Schwinger spinor of the spin-3/2 particle is chosen as $\overline{u}_{\alpha} (p,\lambda') u^{\alpha} (p,\lambda) = - 2M \delta_{\lambda' \lambda}$.
Note that the matrix elements as well as GPDs are defined flavor by flavor.
We omit the flavor index in $H$, $\tilde{H}$ for simplicity and we should multiply the electric or weak charges and sum over flavors to get the conventional form factors. It should be mentioned that 
the time reversal leads to the following constraints, 
\begin{equation}\label{time}
	\begin{aligned}
			H_i (x,\xi,t)=&H_i (x,-\xi,t) 
			\quad \text{with}\quad i=1,2,4,5,6,8,\\
			H_i (x,\xi,t)=& - H_i (x,-\xi,t) 
			\quad \text{with}\quad i=3,7,\\
			\tilde{H}_j (x,\xi,t)=&- \tilde{H}_j (x,-\xi,t) 
			\quad \text{with} \quad j=1,2,3,4,\\
			\tilde{H}_j (x,\xi,t)=& \tilde{H}_j (x,-\xi,t) 
			\quad \text{with} \quad j=5,6,7,8.
	\end{aligned}
\end{equation}\\

Moreover, in the forward limit, the structure functions in deep inelastic scattering at leading twist and leading order in $\alpha_s$ can be expressed in terms of GPDs defined in Eqs.~\eqref{unGPDDefine} and \eqref{poGPDDefine}.
The hardon structure functions are given as
\begin{equation}\label{structurefunctions}
F_1(x) = \sum_{q} e_q^2 F_1^q (x)  \ , \quad %\nonumber \\
b_1(x) = \sum_{q} e_q^2 b_1^q (x) \ , \quad
g_1(x) = \sum_{q} e_q^2 g_1^q (x)  \ , \quad
g_2(x) = \sum_{q} e_q^2 g_2^q (x)  \ ,  
\end{equation}
where the single flavor structure functions are \cite{Jaffe:1988up}
\begin{equation}\label{structurefunctions1}
	\begin{split}
		F_1^q (x) =&\, H_1 (x,0,0) =
		\frac{q^{\frac{3}{2}}_{\uparrow} (x) 
		+ q^{- \frac{3}{2}}_{\uparrow} (x) +
		q^{\frac{1}{2}}_{\uparrow} (x) 
		+ q^{- \frac{1}{2}}_{\uparrow} (x)}{2},\\
		b_1^q (x) =&\, H_4 (x,0,0) = \frac{\left(
		q^{\frac{3}{2}}_{\uparrow} (x) + 
		q^{- \frac{3}{2}}_{\uparrow} (x) \right)
		- \left( q^{\frac{1}{2}}_{\uparrow} (x) 
		+ q^{- \frac{1}{2}}_{\uparrow} (x)\right)}{2},\\
		g_1^q (x) =&\, \tilde{H}_5 (x,0,0) = \frac{3
		\left( q^{\frac{3}{2}}_{\uparrow} (x) 
		- q^{- \frac{3}{2}}_{\uparrow} (x)
		\right) + \left( q^{\frac{1}{2}}_{\uparrow} (x) - q^{-
		\frac{1}{2}}_{\uparrow} (x) \right)}{\sqrt{20}},\\
		g_2^q (x) =&\, \tilde{H}_8(x,0,0)
		= \frac{\left( q^{\frac{3}{2}}_{\uparrow} (x) 
		-q^{- \frac{3}{2}}_{\uparrow} (x) \right) 
		- 3 \left(q^{\frac{1}{2}}_{\uparrow} (x) 
		- q^{- \frac{1}{2}}_{\uparrow} (x) \right)}{\sqrt{20}}.
	\end{split}
\end{equation}

The decomposition of the matrix elements of the vector~\cite{Nozawa:1990gt,Fu:2022bpf} and axial vector~\cite{Alexandrou:2010tj,Alexandrou:2013opa,Jun:2020lfx,Fu:2022bpf} currents for the spin-3/2 case are  
\begin{equation}
    \begin{aligned}
      \langle p',\lambda ' \left| \Bar{\psi}(0) \gamma^\mu 
      \psi (0) \right| p, \lambda \rangle
      = & -2\overline{u}_{\alpha'} (p', \lambda') 
      \left[ g^{\alpha' \alpha} \left( G_1 (t) \frac{P^{\mu}}{M} 
      + G_5 (t) \gamma^{\mu} \right) \right. \\
     & \left. + \frac{P^{\alpha'} P^{\alpha}}{M^2} 
     \left( G_2 (t) \frac{P^{\mu}}{M} + G_6 (t) 
     \gamma^{\mu} \right) \right] u_{\alpha} (p, \lambda),
    \end{aligned}\label{vectord}
\end{equation}
\begin{equation}
    \begin{aligned}
      \langle p',\lambda ' \left| \Bar{\psi}(0) 
      \gamma^\mu \gamma^5 \psi (0) \right| p, \lambda \rangle
     = & -2\overline{u}_{\alpha'} (p', \lambda') 
     \left[ g^{\alpha' \alpha} \left(- \tilde{G}_1 (t) \frac{q^{\mu}}{2M} 
      + \tilde{G}_5 (t) \gamma^{\mu} \right) \right. \\
     & \left. + \frac{P^{\alpha'} P^{\alpha}}{M^2} 
     \left(- \tilde{G}_2 (t) \frac{q^{\mu}}{2M} + \tilde{G}_6 (t) 
     \gamma^{\mu} \right) \right] \gamma^5 u_{\alpha} (p, \lambda),
    \end{aligned}\label{axiald}
\end{equation}
where the different parameters, $ 2 \left( \tilde{G}_1, \tilde{G}_2, \tilde{G}_5, \tilde{G}_6 \right) = \left( -g_3, h_3, g_1, -h_1 \right)$~\cite{Jun:2020lfx}, are used. 
Further, the EMFFs including charge monopole $G_{E0}$, magnetic dipole $G_{M1}$, charge quadrupole $G_{E2}$ and magnetic octupole $G_{M3}$ can be expressed as the linear combinations of the coefficient functions $G_{1,2,5,6}$~\cite{Nozawa:1990gt,Fu:2022rkn}.
Here we employ the same conventions as employed in Ref.~\cite{Fu:2022bpf},
\begin{equation} \label{coeff_E}
	\begin{split}
		E_1 = & H_1 + H_4 - H_5, \quad E_4 = 3 H_4 - 3 H_8, \quad E_8 = 3 H_8,\\
		E_i = & H_i \quad \text{with} \quad i = 2,3,5,6,7,
\end{split}
\end{equation}
and
\begin{equation}\label{coeff_Et}
\begin{split}
	\tilde{E}_5=&\frac{3}{\sqrt{5}} \tilde{H}_5 + \frac{1}{\sqrt{5}}\tilde{H}_8, \quad \tilde{E}_6=\frac{3}{\sqrt{5}} \tilde{H}_6, \quad \tilde{E}_8=\sqrt{5} \tilde{H}_8,\\
	\quad \tilde{E}_j =& \tilde{H}_j \quad \text{with} \quad j = 1,2,3,4,7.
\end{split}
\end{equation}
Based on the Mellin-moments \cite{Diehl:2003ny}, the sum rules connecting GPDs with the coefficient functions and axial vector form factors are
\begin{equation}\label{sumrules}
    \begin{aligned}
    \int ^1_{-1}\text{d} x \, E_i (x,\xi,t)=& G_i(t) \quad \text{with} 
      \quad i=1,2,5,6,\\
    \int ^1_{-1} \text{d} x \, \tilde{E}_i (x,\xi,t)=& \xi \tilde{G}_i(t) 
      \quad \text{with} \quad i=1,2,\\
     \int ^1_{-1} \text{d} x \, \tilde{E}_i (x,\xi,t)=& \tilde{G}_i(t) 
      \quad \text{with} \quad i=5,6,\\
     \int ^1_{-1}\text{d} x \, E_j (x,\xi,t)= \int ^1_{-1}\text{d} x & \, \tilde{E}_j (x,\xi,t) = 0 \quad \text{with} \quad j=3,4,7,8.
    \end{aligned}
\end{equation}
Analogously, the decomposition of the energy-momentum tensor current matrix element \cite{Kim:2020lrs,Fu:2022bpf} is written in terms of GFFs as
\begin{equation}
	\begin{aligned}
	  & \left\langle p', \lambda' \left| \hat{T}^{\mu \nu} (0) \right| p, \lambda \right\rangle\\
	  = & - \overline{u}_{\alpha'} (p', \lambda')
	  \left[ \frac{P^{\mu} P^{\nu} }{M}
	  \left( g^{\alpha' \alpha} F^T_{1, 0} (t) + \frac{2 P^{\alpha'}
	  P^{\alpha}}{M^2} F_{1, 1}^T (t) \right) \right.\\
	  & + \frac{(q^{\mu} q^{\nu} - g^{\mu \nu} q^2)}{4 M}  
	  \left( g^{\alpha'
	  \alpha} F^T_{2, 0} (t) + \frac{2 P^{\alpha'} 
	  P^{\alpha}}{M^2} F_{2, 1}^T(t) \right)\\
	  & + M g^{\mu \nu} \left( g^{\alpha' \alpha} F^T_{3, 0} (t) 
	  + \frac{2 P^{\alpha'} P^{\alpha}}{M^2} F_{3, 1}^T (t) 
	  \right) + \frac{P^{\{ \mu
	  \nobracket} i \sigma^{\nobracket \nu \} q}}{2 M}  
	  \left( g^{\alpha' \alpha}
	  F^T_{4, 0} (t) + \frac{2 P^{\alpha'} P^{\alpha}}{M^2} 
	  F_{4, 1}^T (t)\right)\\
	  & \left. - \frac{1}{M} \left(2 q^{ \{ \mu \nobracket} 
	  g^{\nobracket \nu \} [ \alpha' \nobracket}
	  P^{\nobracket \alpha ]} + 8 g^{\mu \nu} P^{\alpha'} 
	  P^{\alpha} - g^{\alpha'
	  \{ \mu \nobracket} g^{\nobracket \nu \} \alpha} q^2 \right) 
	  F^T_{5, 0} (t) + M g^{\alpha' \{ \mu \nobracket} 
	  g^{\nobracket \nu \} \alpha} 
	  F^T_{6, 0}(t) \right]u_{\alpha} (p, \lambda).
	\end{aligned} \label{GFFs}
\end{equation}
Finally, the sum rules connecting GPDs with GFFs are given as
\begin{equation}
	\begin{aligned}
	   \int ^1_{-1}\text{d} x \, x E_1 (x, \xi, t) =&  F_{1, 0}^T (t) 
	  +  \xi^2 F_{2, 0}^T (t) - 2F_{4, 0}^T (t),\\
	   \int ^1_{-1}\text{d} x \, x E_2 (x, \xi, t) =& 2 F_{1, 1}^T (t) 
	  + 2 \xi^2 F_{2, 1}^T(t) - 4F_{4, 1}^T (t),\\
	   \int ^1_{-1}\text{d} x \,x E_3 (x, \xi, t) =& 8 \xi F_{5, 0}^T (t),\\
	   \int ^1_{-1}\text{d} x \,x E_4 (x, \xi, t) =& \frac{2 t}{M^2} F^T_{5, 0} (t) + 2 F^T_{6, 0} (t),\\
	   \int ^1_{-1}\text{d} x \,x E_5 (x, \xi, t) =& 2 F^T_{4, 0} (t),\\ 
	   \int ^1_{-1}\text{d} x \,x E_6 (x, \xi, t) =& 4 F_{4, 1}^T (t),\\
	   \int ^1_{-1}\text{d} x \,x E_i (x, \xi, t) =& 0, \quad \tmop{with} \quad i = 7, 8.
	\end{aligned} \label{GFFs0}
\end{equation}

\section{Diquark spectator approach}
\label{model}
\quad\quad In the present work, we consider the $\Delta^+$ as an attempt to characterize the multidimensional structure of the spin-3/2 particles. 
In the picture of the quark model, the $\Delta^+$ isobar is composed of three light quarks, two $u$ quarks and one $d$ quark. 
The quantum numbers of  $\Delta^+$ are $I (J^P)=3/2 (3/2^+)$ and it requires that both the isospin and spin of each pair of quarks be 1. 
Therefore it is convenient to regard two quarks in $\Delta^+$ as a whole, i.e. diquark.
Certainly, the electric charge of different quarks and diquarks needs to be considered carefully.
However, the calculations of the diquark GPDs are difficult because of the more complex integral, so we calculate the GPDs for each flavor quark using the diquark spectator approach instead of the quark-diquark approach and sum all the quark contributions.
Need to note that here is different from our previous work \cite{Fu:2022rkn}.
As a check of consistency of our calculation, the EMFFs and GFFs, which are obtained by direct calculation in Ref.~\cite{Fu:2022rkn}, are compared with those given by the GPDs method through the sum rules, and the results are identical. In contrast to the Ref.~\cite{Fu:2022rkn}, the diquark contribution is not calculated in the present work and it is approximated by the two times of the quark one.  

Note that the GPDs and the form factors can be calculated in different reference frames \cite{Belitsky:2005qn} and the Breit frame, $q^+ = -q^-$, is employed for convenience in this work, where the initial and final momenta are
\begin{equation}
	\begin{split}
		p=(p^0-\frac{q_z}{2}, p^0+\frac{q_z}{2}, -\frac{\bm{q}_\perp}{2}), \quad
		p'=(p^0+\frac{q_z}{2}, p^0-\frac{q_z}{2}, \frac{\bm{q}_\perp}{2}).
	\end{split}
\end{equation}

\begin{figure}[h]
	\centering
	\includegraphics[height=3cm]{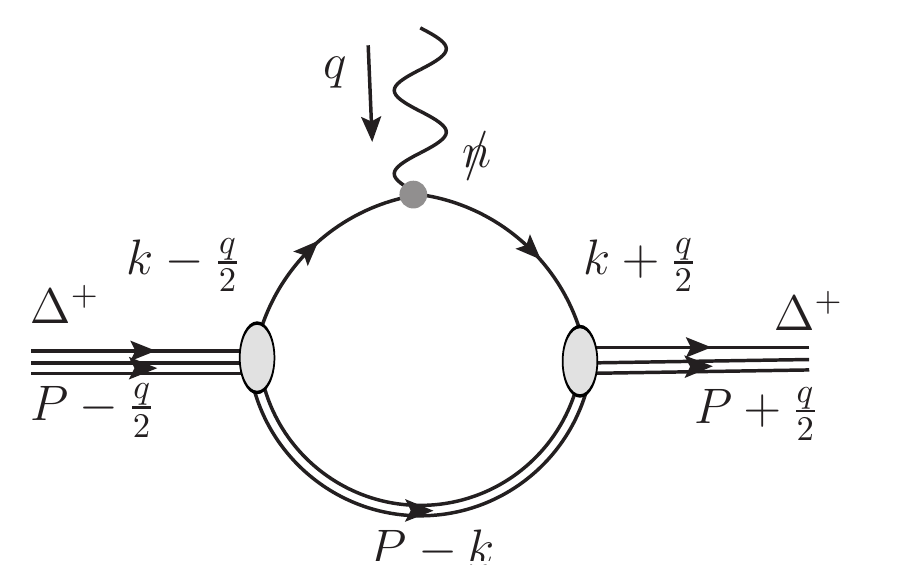}
	\caption{\small{Feynman diagram for the $\Delta^+$ GPDs using the diquark spectator approach, and the single (double) line stands for the quark (diquark).}}
	\label{feynmandiagram}
\end{figure}

Figure~\ref{feynmandiagram} is the Feynman diagram of the diquark spectator approach.
In this work, the high order term $k^\mu k^\nu /m_V^2$ in the vector propagator is neglected in order to simplify the calculation and to give the finite results~\cite{Dong:2009yp}.
To calculate the unpolarized GPDs defined in Eq.~\eqref{unGPDDefine} using the analogous method to Ref.~\cite{Fu:2022rkn}, one can get
\begin{equation}\label{unint1}
	\mathcal{H}^{\alpha' \alpha}
	  = -\frac{i}{2} c_1^2 \int \frac{\text{d}^2 k_\perp \text{d} k^+ \text{d} k^-}{(2 \pi)^4}\frac{\delta(k\cdot n - x P \cdot n)}{\mathfrak{D}}
	  \Gamma^{\alpha' \beta'} \left(\slashed{k}+\frac{\slashed{q}}{2}+m_q \right)
	  g_{\beta' \beta} \slashed{n} \left(\slashed{k}-\frac{\slashed{q}}{2}
	  +m_q \right) \Gamma^{\beta \alpha},
\end{equation}
where
\begin{equation}\label{OmegaD}
    \begin{split}
    \mathfrak{D}=& \left[\left(k+ \frac{q}{2} \right)^2 - m_q^2+i \epsilon\right]
    \left[\left(k- \frac{q}{2} \right)^2 - m_q^2+i \epsilon\right]
	\left[\left(k-P\right)^2-m_D^2+i \epsilon \right] \left[\left( k-P \right) ^2 -m_R^2+i \epsilon \right]^2\\
    & \left[\left( k+\frac{q}{2} \right) ^2 -m_R^2+i \epsilon\right]
	\left[\left( k-\frac{q}{2} \right) ^2 -m_R^2+i \epsilon\right],
    \end{split}
\end{equation}
and the effective form of the vertex function is \cite{Scadron:1968zz}
\begin{equation}\label{vertexfunction}
		\Gamma^{\alpha \beta} = \left[ g^{\alpha \beta} + c_2 \gamma^\beta \Lambda^\alpha + c_3 \Lambda^\beta \Lambda^\alpha\right],
\end{equation}
where $\Lambda$ is the relative momentum between the spin-1/2 and spin-1 partons. In Eq.~\eqref{unint1}, we also employ the scalar function of the loop momentum, $\Xi (p_1,p_2,m_R)=c_1/(p_1^2-m_R^2+i \epsilon)(p_2^2-m_R^2+i \epsilon)$, to perform the regularization. Although this regularization method breaks the gauge invariance compared with some other methods, such as Pauli-Villars regularization \cite{Pauli:1949zm}, it simplifies the numerical calculation.
The couplings parameters $c_1$, $c_2$, and $c_3$ and the cut-off parameter $m_R$ are determined as follows: $c_1$ is fixed by the normalization of the charge, and $c_2$, $c_3$, and $m_R$ are obtained by fitting the results of EMFFs of the lattice QCD calculation.

To extract GPDs from the $\mathcal{H}^{\alpha' \alpha}$, one needs some identities and on-shell identities like Schouten identity and Gordon identity which are listed in \ref{appendixuseful}. 
To carry out the integrals involving $k^\mu$, $k^\mu k^\nu$ and $k^\mu k^\nu k^\rho$ terms, we applied some transformations of these integrals as shown in \ref{appendixtransformation}.
Now for each of the GPDs, the integrand only contains the scalar product of $k$, and they will be calculated by contour integral~\cite{Sun:2017gtz}.
Analogously, the polarized GPDs can be given using the same procedure and we just replace $\slashed{n}$ by $\slashed{n} \gamma^5$ in Eq.~\eqref{unint1}. Some useful identities for the polarized case are also listed in \ref{appendixuseful}.

\section{Numerical results}\label{result}

\subsection{Results for GPDs}\label{result_GPD}

In this present work, we use the parameters, $\Delta$ resonance mass $M$, quark mass $m_q$, diquark mass $m_D$, cut-off mass $m_R$ and $c_2$, $c_3$ in the vertex~\eqref{vertexfunction}, same as Ref.~\cite{Fu:2022rkn} which are listed in Table \ref{parameters}.

\begin{table}[ht]
    \centering
	\begin{tabular}{  c  c  c  c  c  c  c }
		\hline
		\hline
		\specialrule{0em}{2pt}{1.5pt}
		$M/\text{GeV}$	& $m_q/\text{GeV}$	& $m_D/\text{GeV}$	& $m_R/\text{GeV}$	& $c_2/\text{GeV}^{-1}$	& $c_3/\text{GeV}^{-2}$\\
		\hline
		\specialrule{0em}{1.5pt}{2pt}
		1.085	& 0.4	& 0.76	& 1.6	& 0.703	& 0.412	\\
		\hline
		\hline
	\end{tabular}
    \caption{\small{The parameters used in this work.}}
    \label{parameters}
\end{table}

\begin{figure}[h]
	\centering
	\includegraphics[height=4cm]{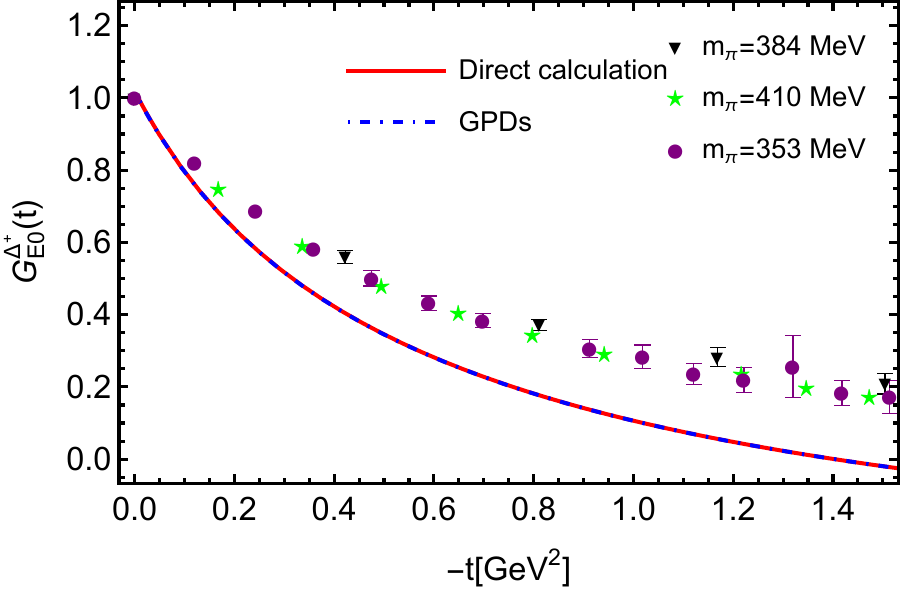} \quad
	\includegraphics[height=4cm]{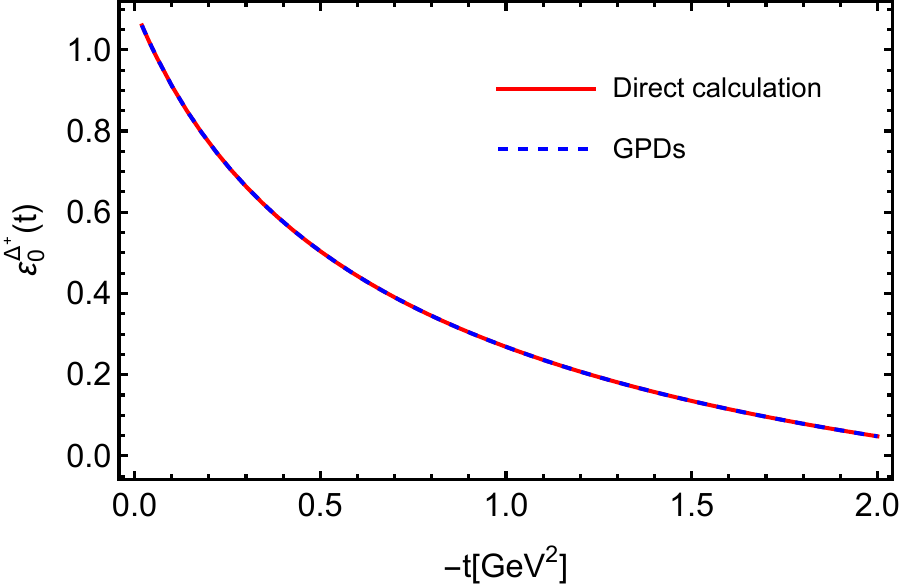}
	\caption{The charge and energy monopole form factors of $\Delta^+$.}
	\label{figureFFs}
\end{figure}

The constraint $\lvert \xi \rvert \leq 1/\sqrt{1-4 M^2/t}$ leads to a lower bound on the value of the squared momentum transfer $t \le -\lvert t \rvert_{\text{min}}$, where $\lvert t \rvert_{\text{min}} = (0,\,0.9)$ $\text{GeV}^2$ when $\xi=(0,\,-0.4)$. 
Before showing the numerical GPDs, we calculate the charge and energy monopole form factors using the sum rules of our obtained GPDs. As shown in Fig.~\ref{figureFFs}, it is not surprising that the results agree with those of our previous work~\cite{Fu:2022rkn}.
Fig.~\ref{figureFFs} also verifies that the diquark spectator approach is a reasonable approximation.

\begin{figure}[h]
	\centering
	\includegraphics[height=5cm]{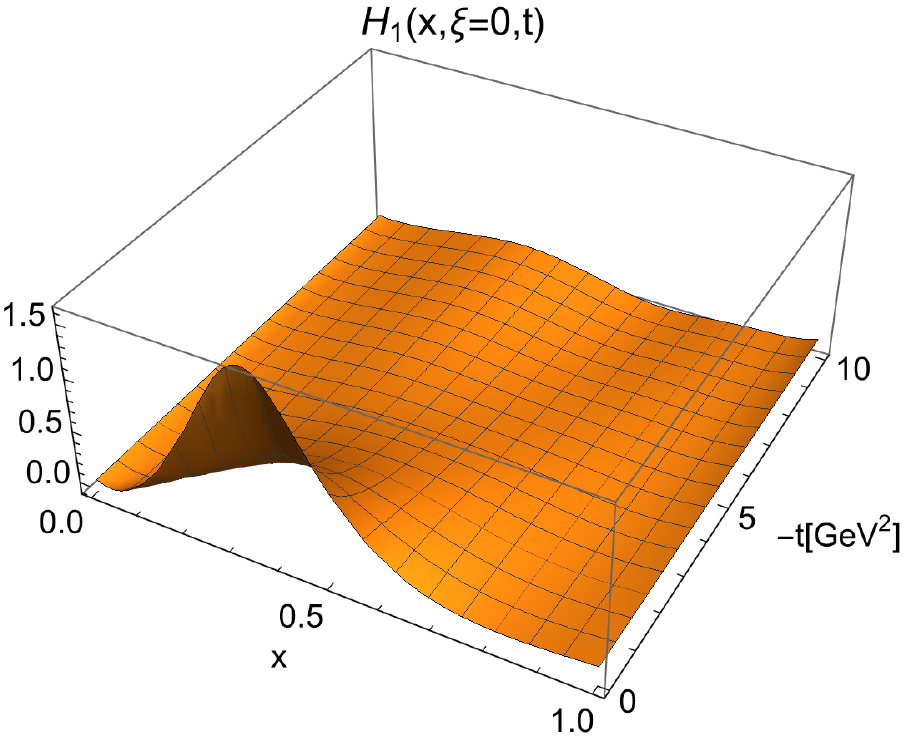} \quad
	\includegraphics[height=5cm]{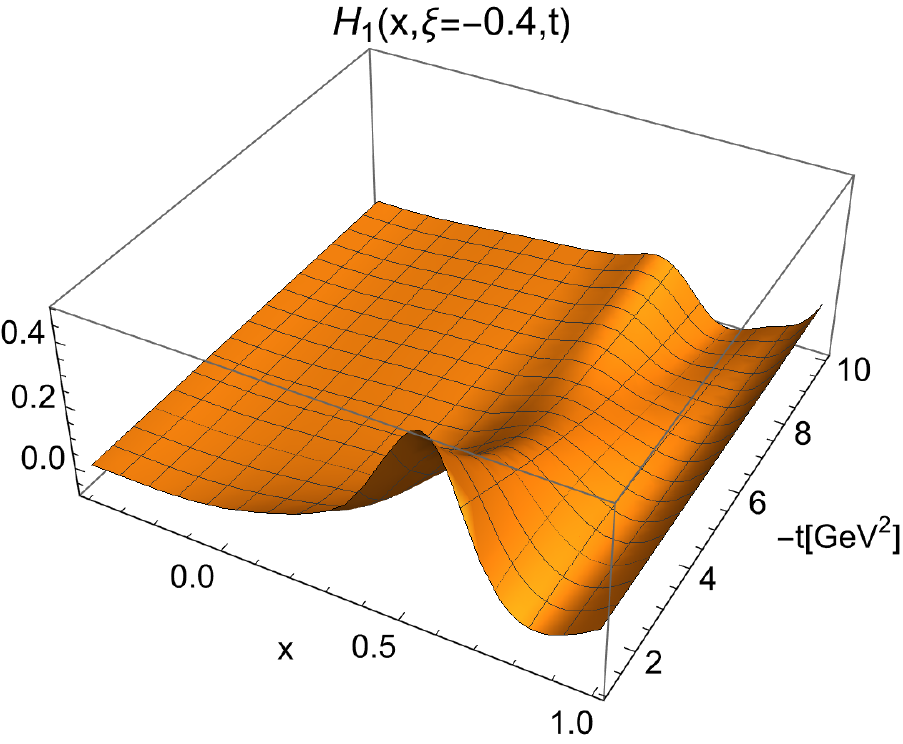}\\
	\includegraphics[height=5cm]{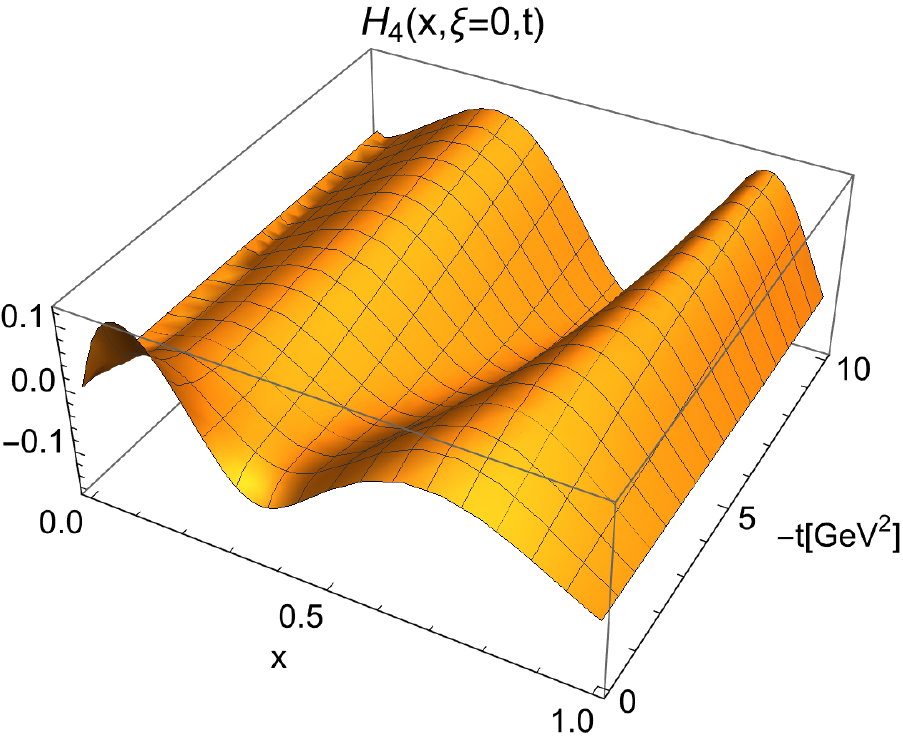} \quad
	\includegraphics[height=5cm]{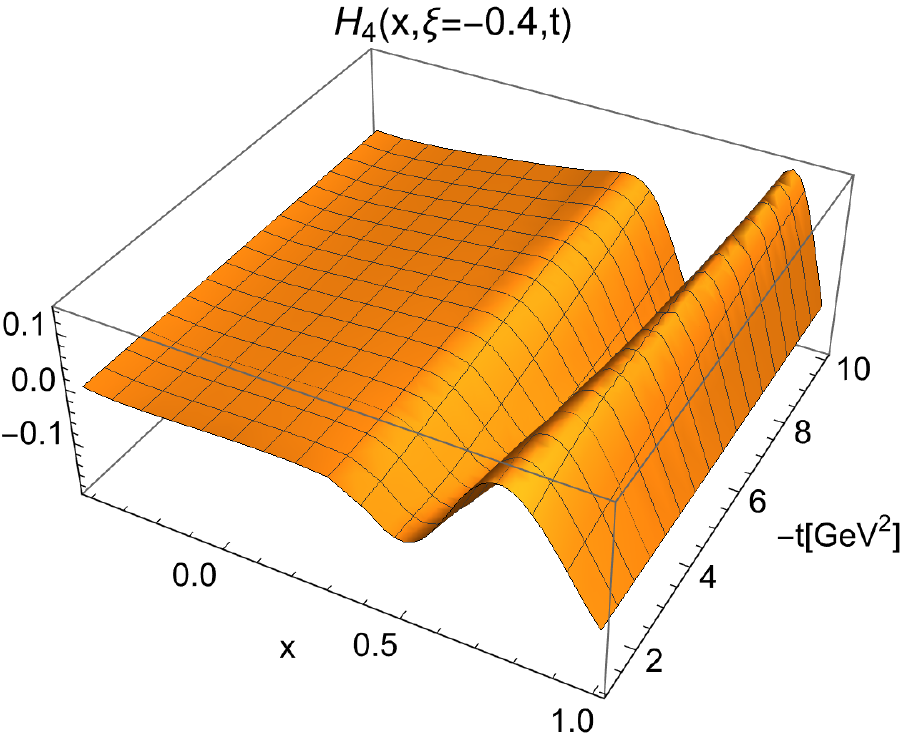}
	\caption{\small{The 3D $d$ quark unpolarized GPDs of $\Delta^+$ $H_1$ and $H_4$ as functions of $x$ and $-t$ at $\xi =0 $ and $\xi = -0.4$.}}
	\label{figureh1}
\end{figure}

\begin{figure}[h]
	\centering
	\includegraphics[height=5cm]{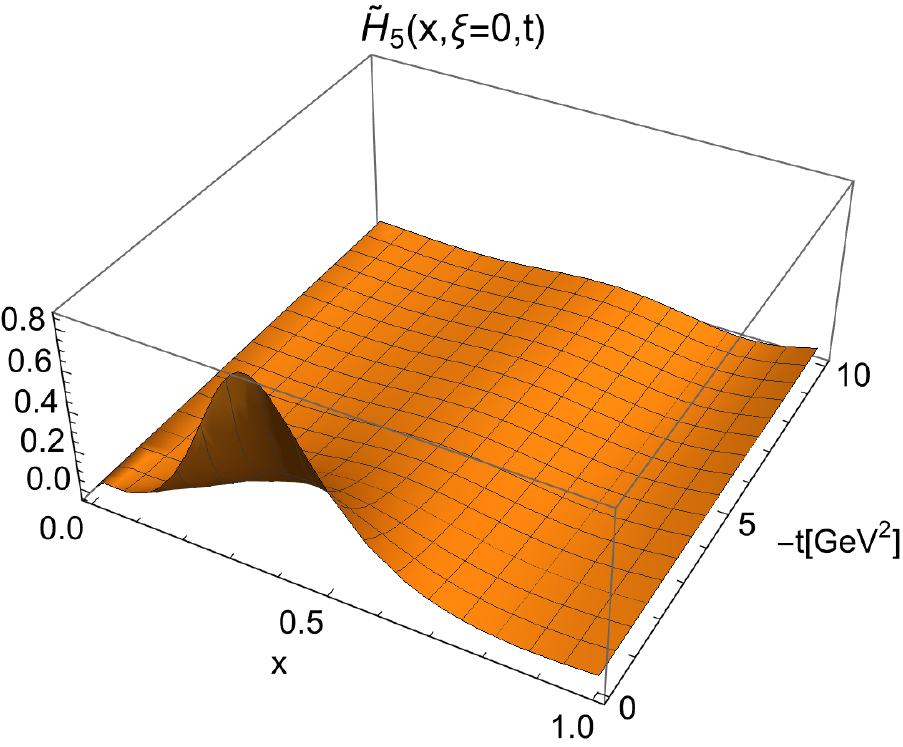} \quad
	\includegraphics[height=5cm]{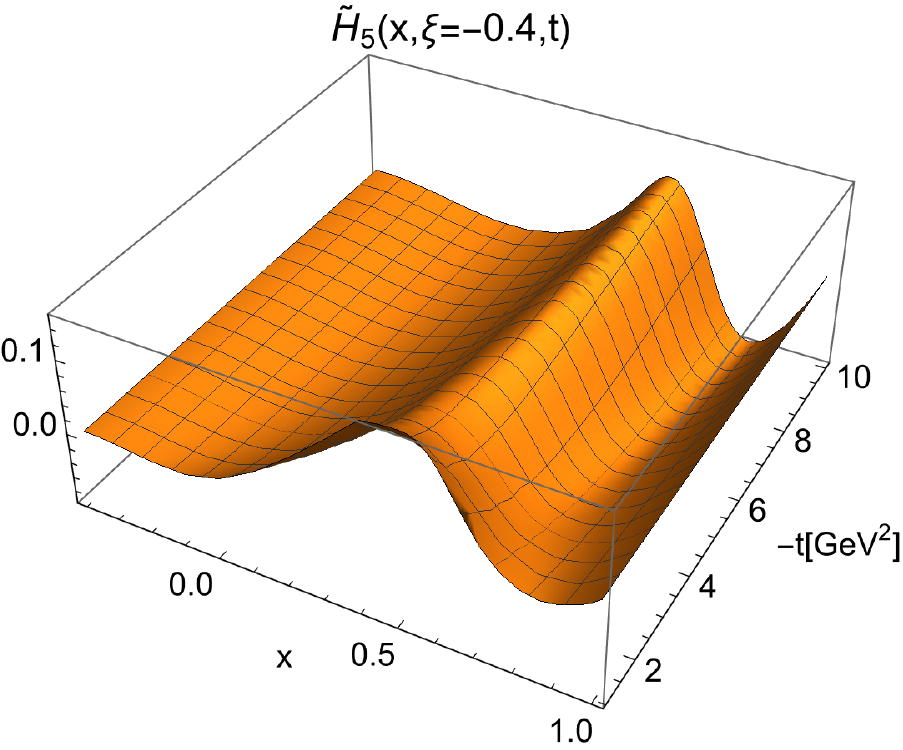} \\
	\includegraphics[height=5cm]{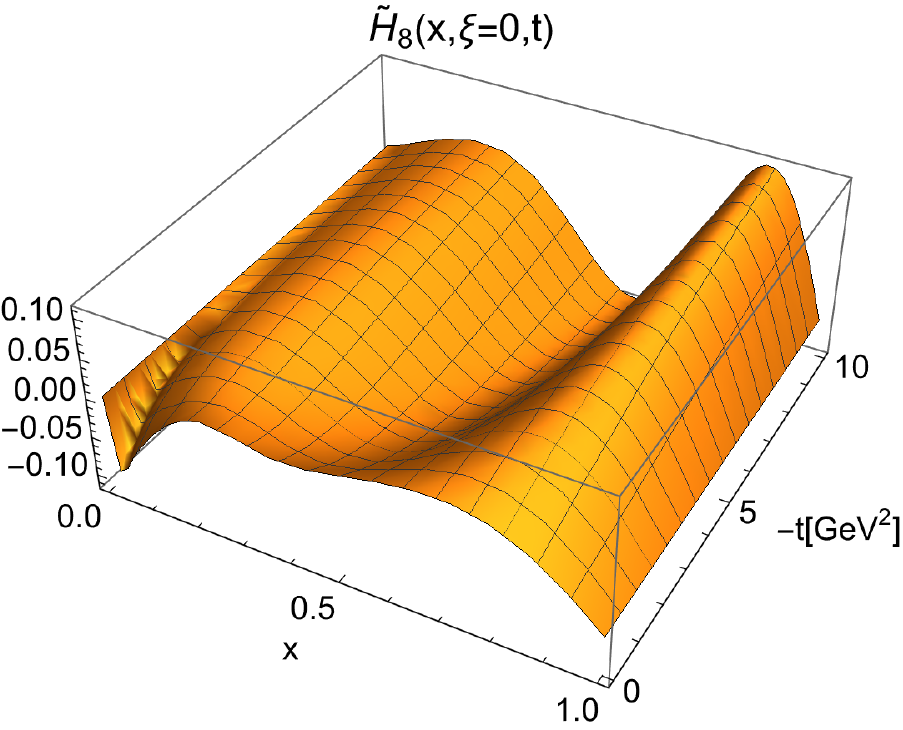} \quad
	\includegraphics[height=5cm]{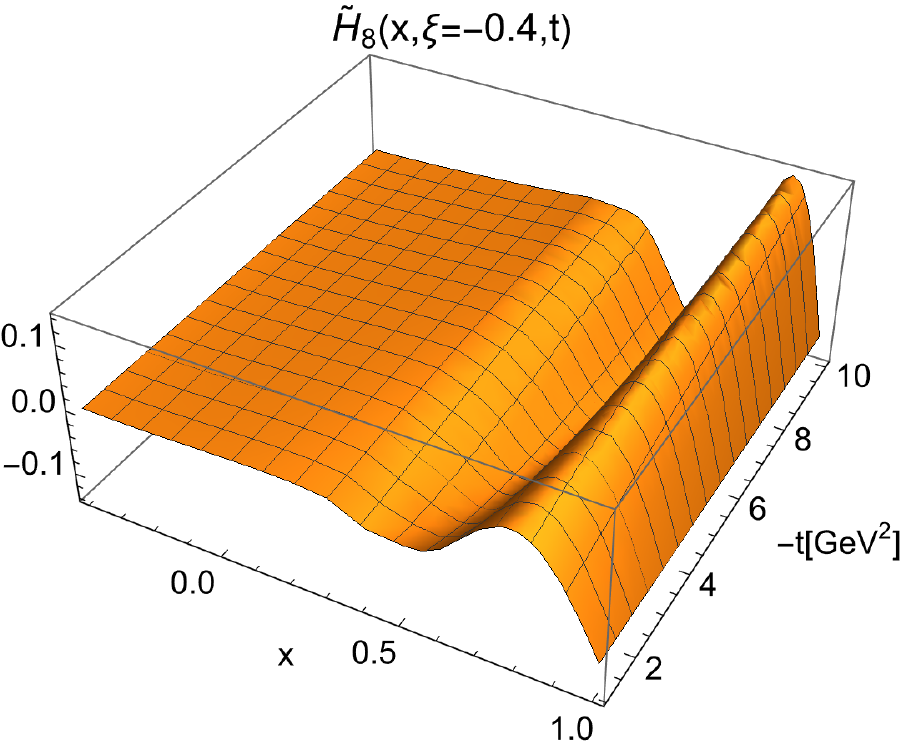}
	\caption{\small{The 3D $d$ quark polarized GPDs of $\Delta^+$ $\tilde{H}_5$ and $\tilde{H}_8$ at $\xi =0 $ and $\xi = -0.4$.}}
	\label{figureh5}
\end{figure}

\begin{figure}[h]
	\centering
	\includegraphics[height=4cm]{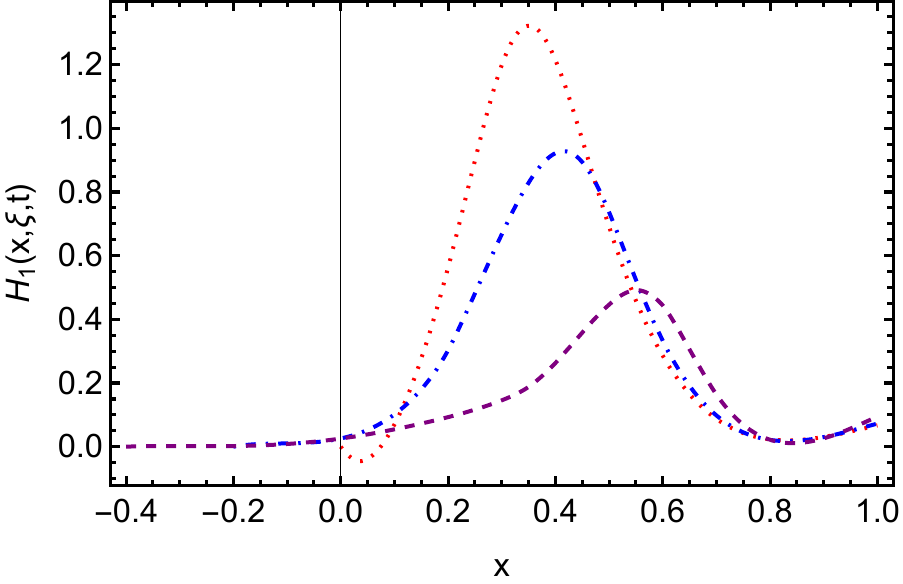} \quad
	\includegraphics[height=4cm]{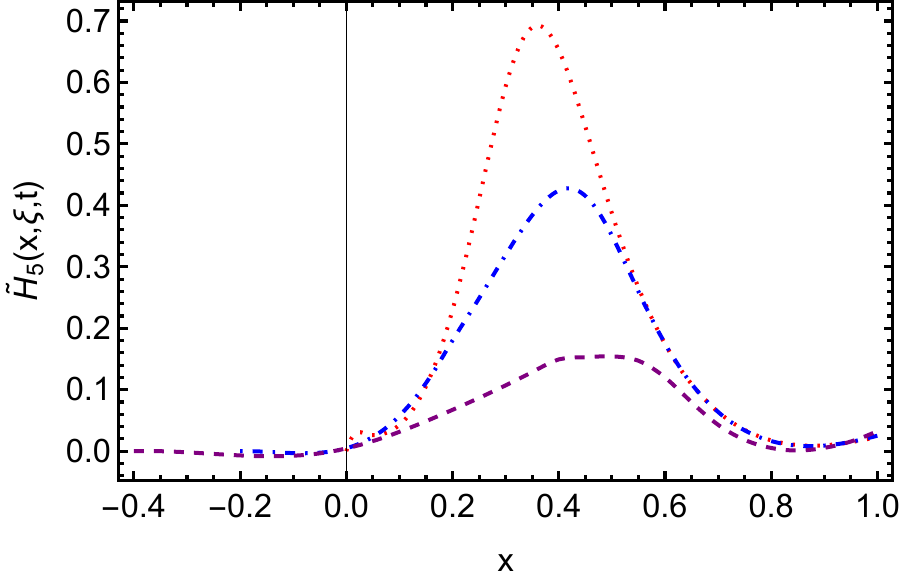}
	\caption{\small{The GPDs $H_1$ and $\tilde{H}_5$ with different $\xi $ and $t$, the red dotted line, blue dot dash line and purple dashed line represent the GPDs with $(\xi,\,-t)=(0,\,0.1~ \text{GeV}^2), \, (-0.2,\, 0.3~ \text{GeV}^2)$ and $(-0.4,\,1.1~ \text{GeV}^2)$ respectively.}}
	\label{figuret}
\end{figure}

The results for the contribution of $d$ quark to the GPDs of $\Delta^+$ will be given as an example without loss of generality since the GPDs of different quarks in all the $\Delta$ isobars are the same in the present model except for the different quark numbers.
Figs.~\ref{figureh1} and \ref{figureh5} show the 3D plots of unpolarized GPDs $H_{1,4}$ and polarized GPDs $\tilde{H}_{5,8}$ as the functions of variables $x$ and $t$ at two different skewnesses $\xi=(0,\,-0.4)$.
Note that the constraints by time reversal, Eq.~\eqref{time}, are satisfied in our numerical results. We find that there is a tendency that the maximums or minimums of GPDs shift to large $x$ as $\lvert \xi \rvert$ increases. This tendency can be elucidated more clearly by the cutting planes, which clearly show how the GPDs $H_1$ (corresponding to the structure function $F_1$ and the charge and energy monopole form factors) and $\tilde{H}_5$ (corresponding to the structure function $g_1$ and the axial vector charge form factor) vary with $\xi$ in Fig.~\ref{figuret}.
Similar to the case of $H_1$ and $\tilde{H}_5$, other GPDs including $H_{2,5,6}$ and $\tilde{H}_{6}$ share the same feature.
Here $\lvert t \rvert_{\text{min}}$ is determined by fixing $\bm{q}_\perp \rightarrow 0$ because of the constraint $\lvert \xi \rvert \leq 1/\sqrt{1-4 M^2/t}$ and $-t$ taken here is larger than $\lvert t \rvert_{\text{min}}$ due to the calculation instability around the boundary.
An interesting observation of Fig.~\ref{figuret} is that: each curve reaches its maximum as the momentum fraction carried by the parton gets close to the mass ratio between the quark and the diquark, i.e. $\frac{x_{\text{max}}+\lvert \xi \rvert}{1-x_{\text{max}}} \sim \frac{m_q}{m_D}$. It gives the position of the maximum as $x_{\text{max}} \sim \frac{m_q+m_D \lvert \xi \rvert}{m_q+m_D}$.

\begin{figure}[h]
	\centering
	\includegraphics[height=4cm]{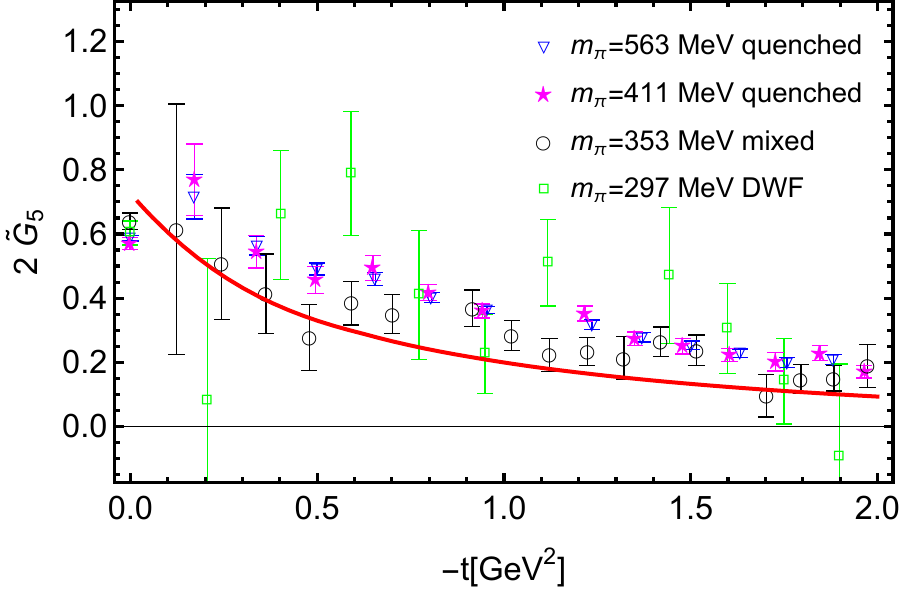} \quad
	\includegraphics[height=4cm]{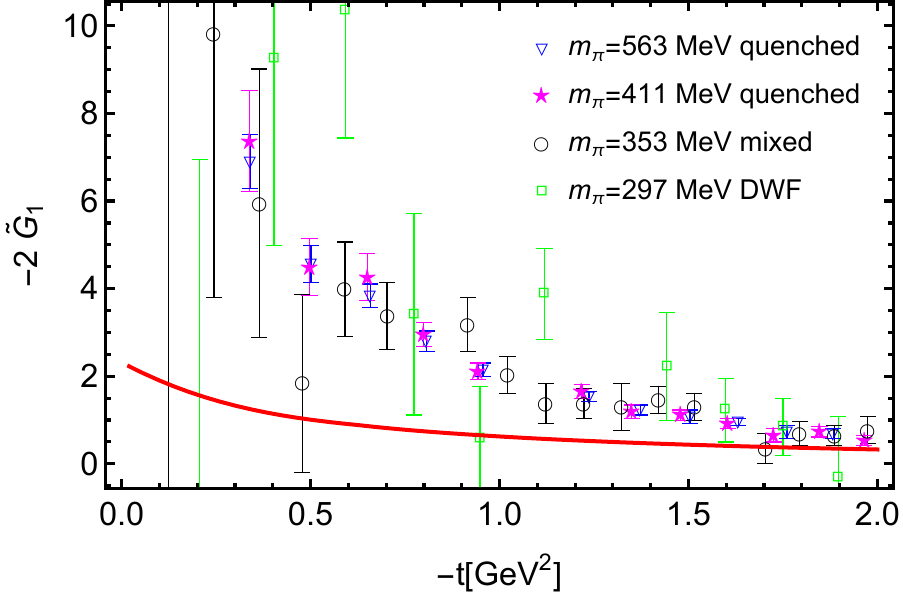} \\
	\includegraphics[height=4cm]{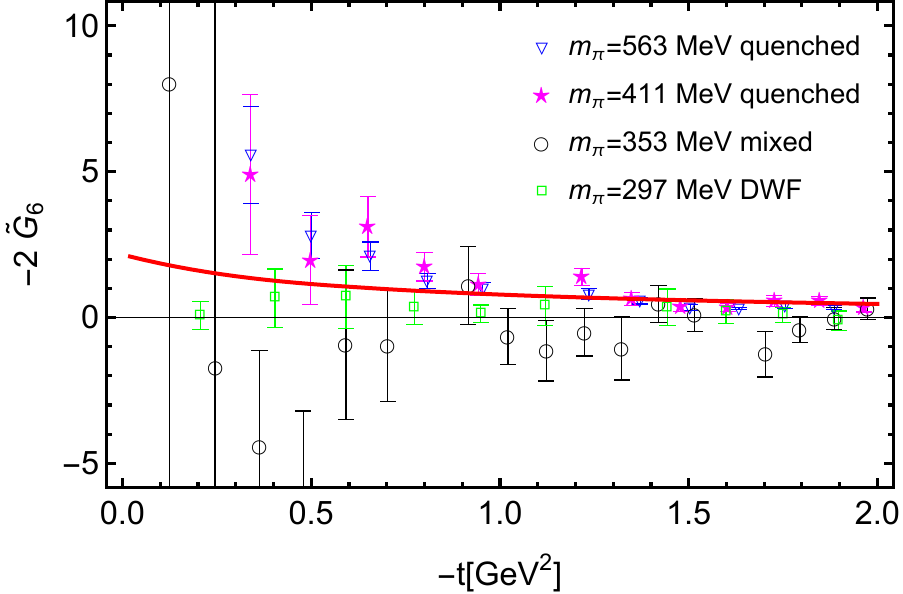} \quad
	\includegraphics[height=4cm]{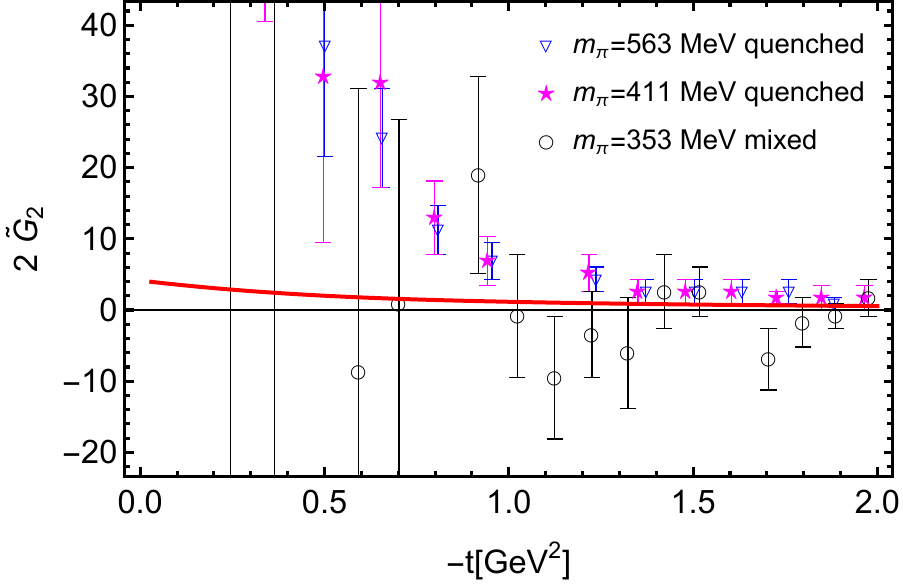}
	\caption{\small{The axial vector form factors of $\Delta^+$ as functions of $-t$ in comparison with lattice QCD results \cite{Alexandrou:2013opa}.}}
	\label{figureAxial}
\end{figure}

The axial vector form factors of $\Delta^+$ are given in Fig.~\ref{figureAxial} from the sum rules of polarized GPDs in Eq.~\eqref{sumrules}. In Fig.~\ref{figureAxial}, the solid lines are our results, and the results from the lattice QCD calculation in Ref.~\cite{Alexandrou:2013opa} are included for comparison.
Our result for the axial charge is $g_A = 2 \tilde{G}_5(0)=0.727$, which is consistent with one in Refs.~\cite{Alexandrou:2013opa}.
In addition, there are also some numerical results on the axial vector form factors of $\Delta^+$ 
with different models \cite{Jun:2020lfx} with which our results of the form factors $\tilde{G_5}$ and $\tilde{G}_6$ are consistent in general.

\begin{figure}[h]
	\centering
	\includegraphics[height=4cm]{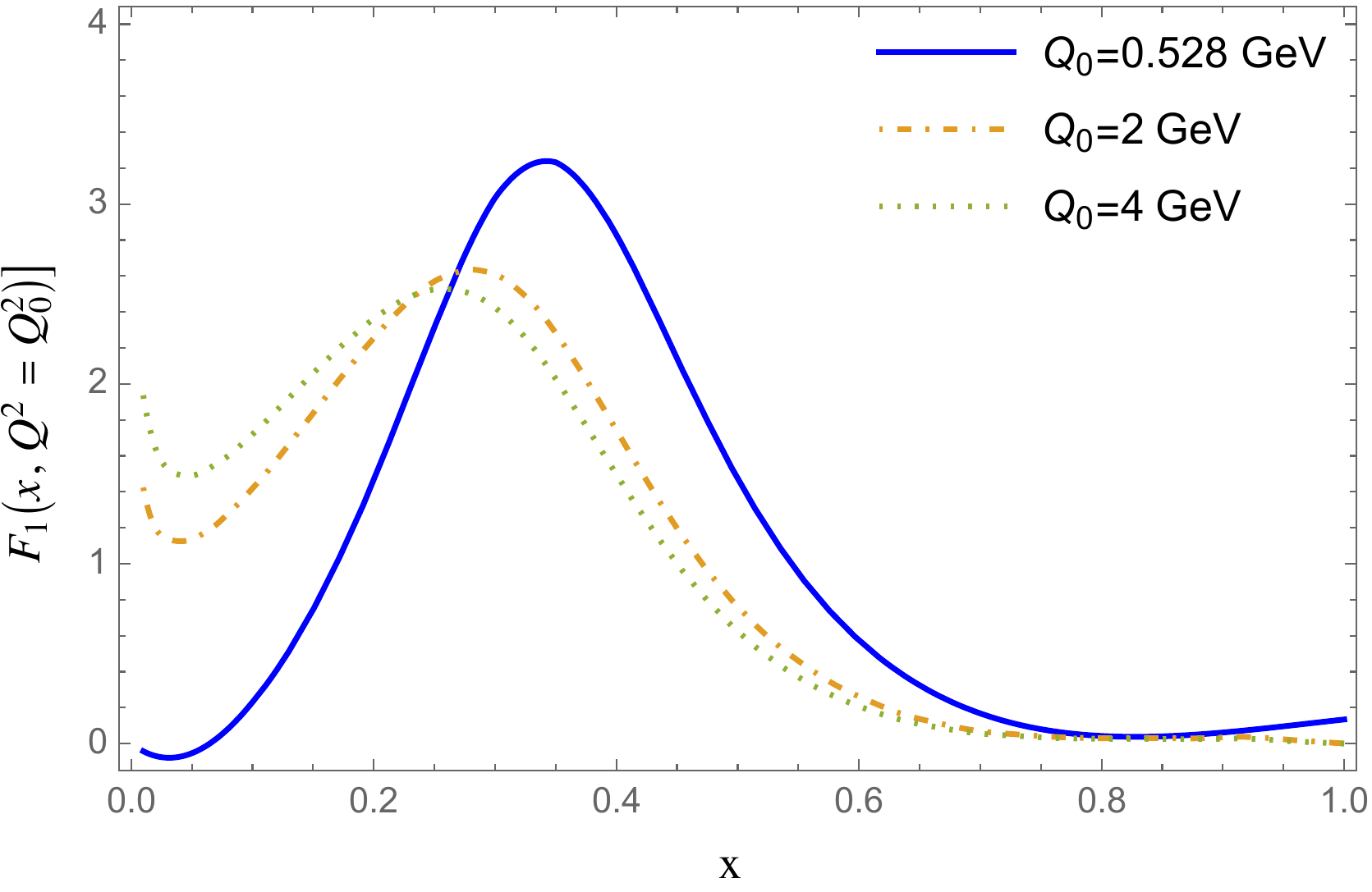} \quad
	\includegraphics[height=4cm]{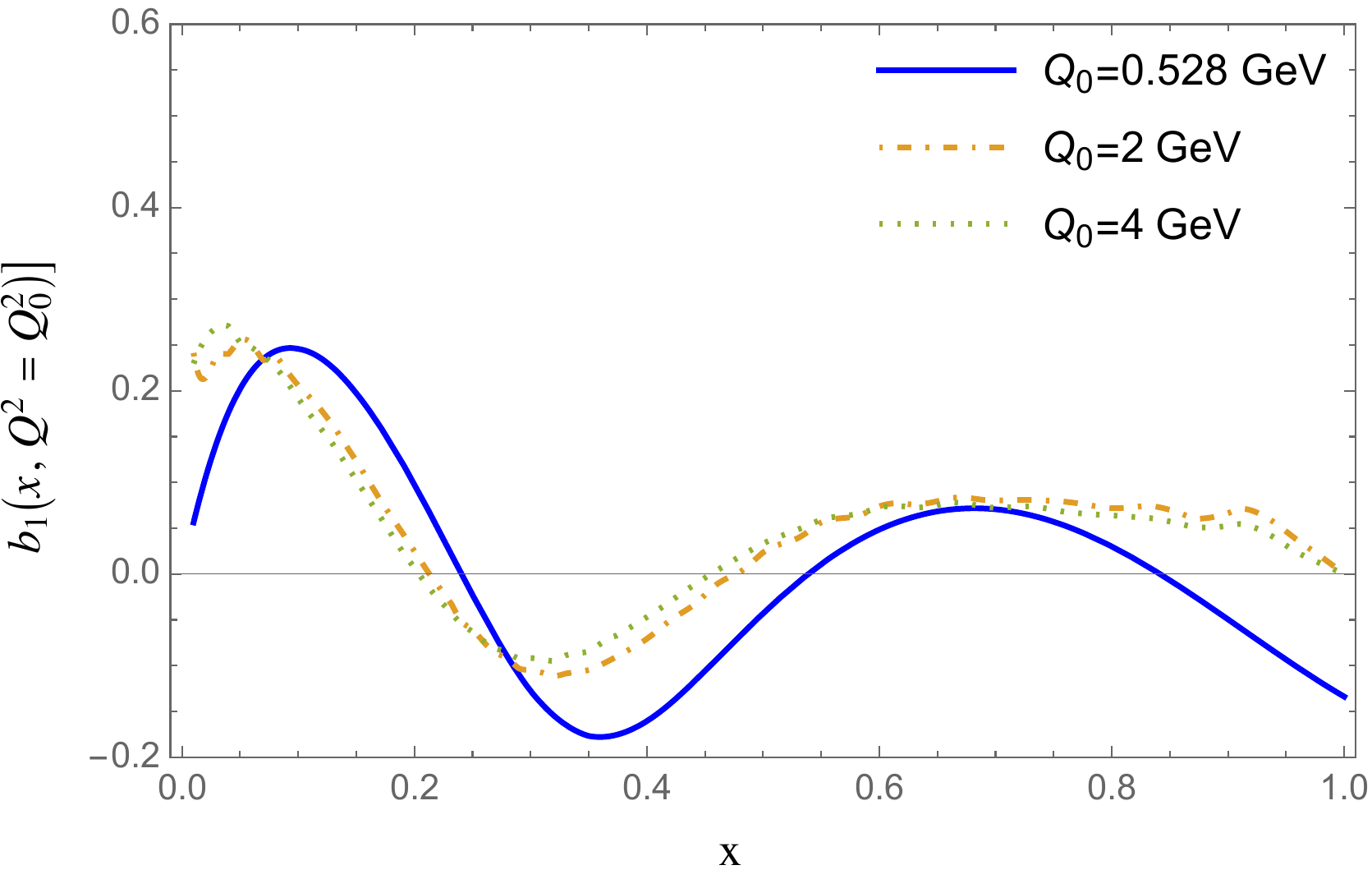} \\
	\includegraphics[height=4cm]{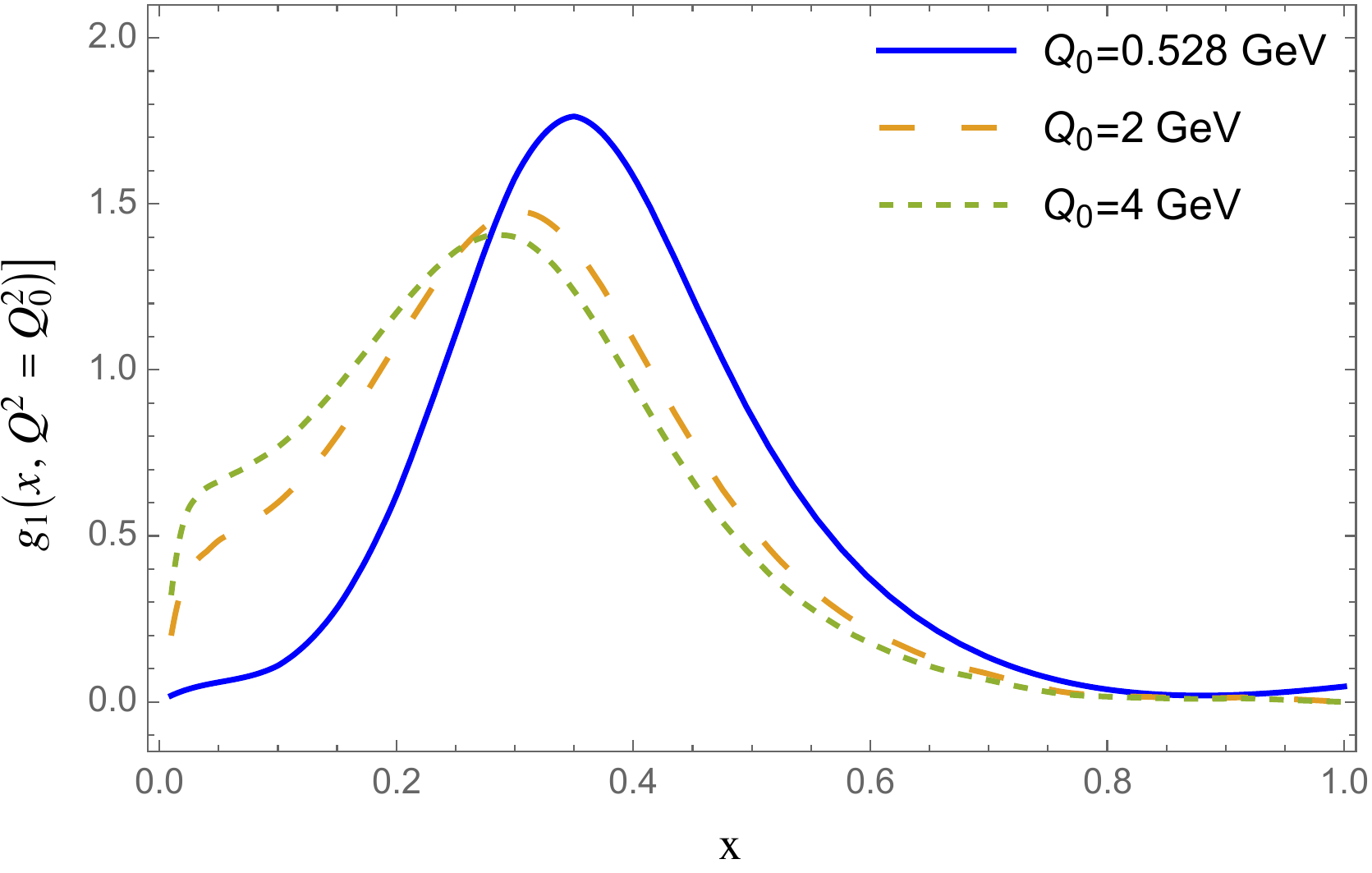} \quad
	\includegraphics[height=4cm]{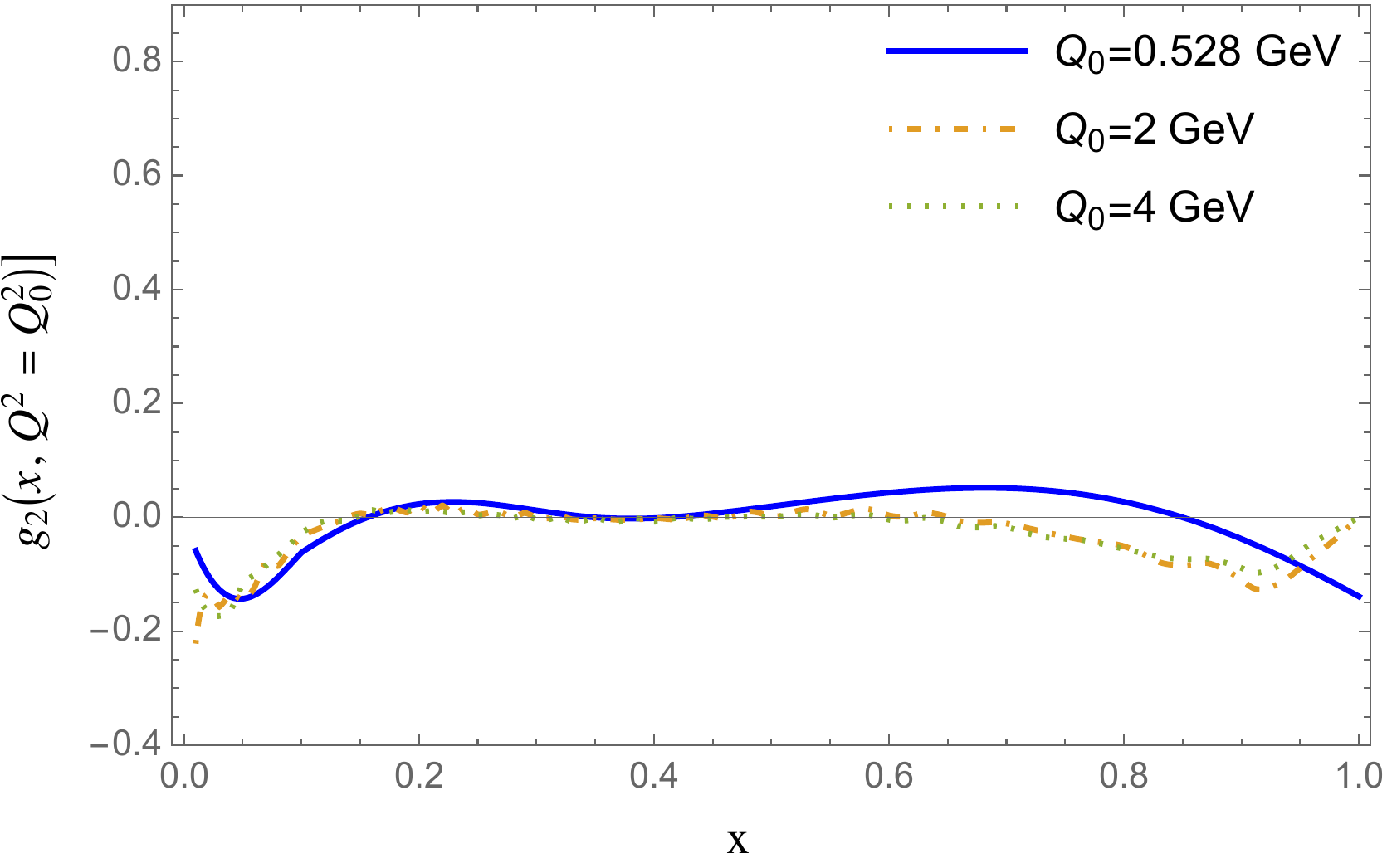}
	\caption{\small{The structure functions of $\Delta^+$ as functions of $x$ at different scales.}}
	\label{figureSF}
\end{figure}

In the forward limit, there are four structure functions as derived in Eqs.~\eqref{structurefunctions} and \eqref{structurefunctions1}. The numerical results for these structure functions of $\Delta^+$ employing the diquark spectator approach are shown in Fig.~\ref{figureSF}, which also include the distributions of structure functions after the DGLAP evolution that will be discussed in the following.
One can find that there are maximums at $x_{\text{max}} \sim m_q/(m_q+m_D)$ in $F_1(x)$ and $g_1(x)$ like the mentioned GPDs and we have been elucidated this character in the above. Notice that the structure functions do not go to zero when $x \rightarrow 1$.
We speculate that the nonzero property occurred due to the effective vertex in Eq.~\eqref{vertexfunction}, and the terms accompanying the parameters $c_2$ and $c_3$ are just simplified forms of some more physical vertex functions like the spin-1 case \cite{Sun:2017gtz}. 
In addition, the structure functions $b_1(x)$ and $g_2(x)$ satisfy the relations,
\begin{equation} \label{eq:b_1_g_2_sum_rule}
    \int ^1_0 \text{d} x \, b_1(x) = 0, \quad
    \int ^1_0 \text{d} x \, g_2(x) = 0,
\end{equation}
based on the definitions of the structure functions \eqref{structurefunctions1} and the sum rules \eqref{sumrules}.

\subsection{Evolution of the structure functions}

The GPDs are scale-dependent quantities. The nonperturbative calculation by low-energy models, like the present quark model, should be treated corresponding to rather low scales with respect to the perturbative QCD. Thus the QCD evolution is needed to compare the predictions of the quark model with experimental data or lattice QCD calculation \cite{Muller:1994ses,Ji:1996nm,Radyushkin:1997ki,Blumlein:1997pi,Golec-Biernat:1998zbo,Kivel:1999wa}. 
In the forward limit, the GPD $H_1(x,0,0)$ corresponds to the usual PDF.
In the quark model, the PDF only is for the valence quarks and its integral over all $x\in [0,1]$ gives 1 which means the total momentum is carried by the valence quarks and no sea quark or gluon contributions.
This condition holds in the so-called quark model scale which is believed to blow 1 GeV~\cite{Sun:2017gtz,Sun:2018ldr}. The experimental data or lattice QCD calculations for lower spin systems like pion and rho mesons are usually given at a scale $Q$ around a few GeV. For the possible experimental or lattice QCD calculations for spin-3/2 systems, one expects that it also has a scale with a similar magnitude.
Therefore, the evolution is required to make predictions for the larger scale from the quark model results. We apply the QCDNUM package~\cite{Botje:2010ay} to perform the DGLAP evolution of the unpolarized structure function $F_1\left(x\right)$ and its $x-$moments at next-to-next-to-leading order (NNLO) and that of polorized structure functions $b_1\left(x\right)$, $g_1\left(x\right)$, and $g_2\left(x\right)$ at next-to-leading order (NLO) in the $\overline{\text{MS}}$ scheme with the number of active quark flavours $n_f=5$.
The $x-$moments, i.e. the so-called Mellin moments, of a function $f(x)$ is defined as
\begin{equation}
M_n(f)=\int_0^1 x^{n-1}f(x)dx \ .
\end{equation}

We adopt the same value for the quark model scale $Q_0=0.528$GeV \cite{Sun:2017gtz} which is obtained by comparing the lattice QCD result for $\rho$ meson with results from a similar quark model. In Fig.~\ref{figureSF}, the evolution of the structure functions is performed up to $Q=4$ GeV. It shows that the distribution in the small $x$ region becomes more dominant for $F_1$ and $g_1$ as the scale increases. For the zero-order moments, the valence quark numbers and the sum rules for structure functions $b_1(x)$ and $g_2(x)$ in Eq.~\eqref{eq:b_1_g_2_sum_rule} hold unchanged within negligible errors during the evolution procedure.

\begin{figure}[h]
	\centering
	\includegraphics[height=4cm]{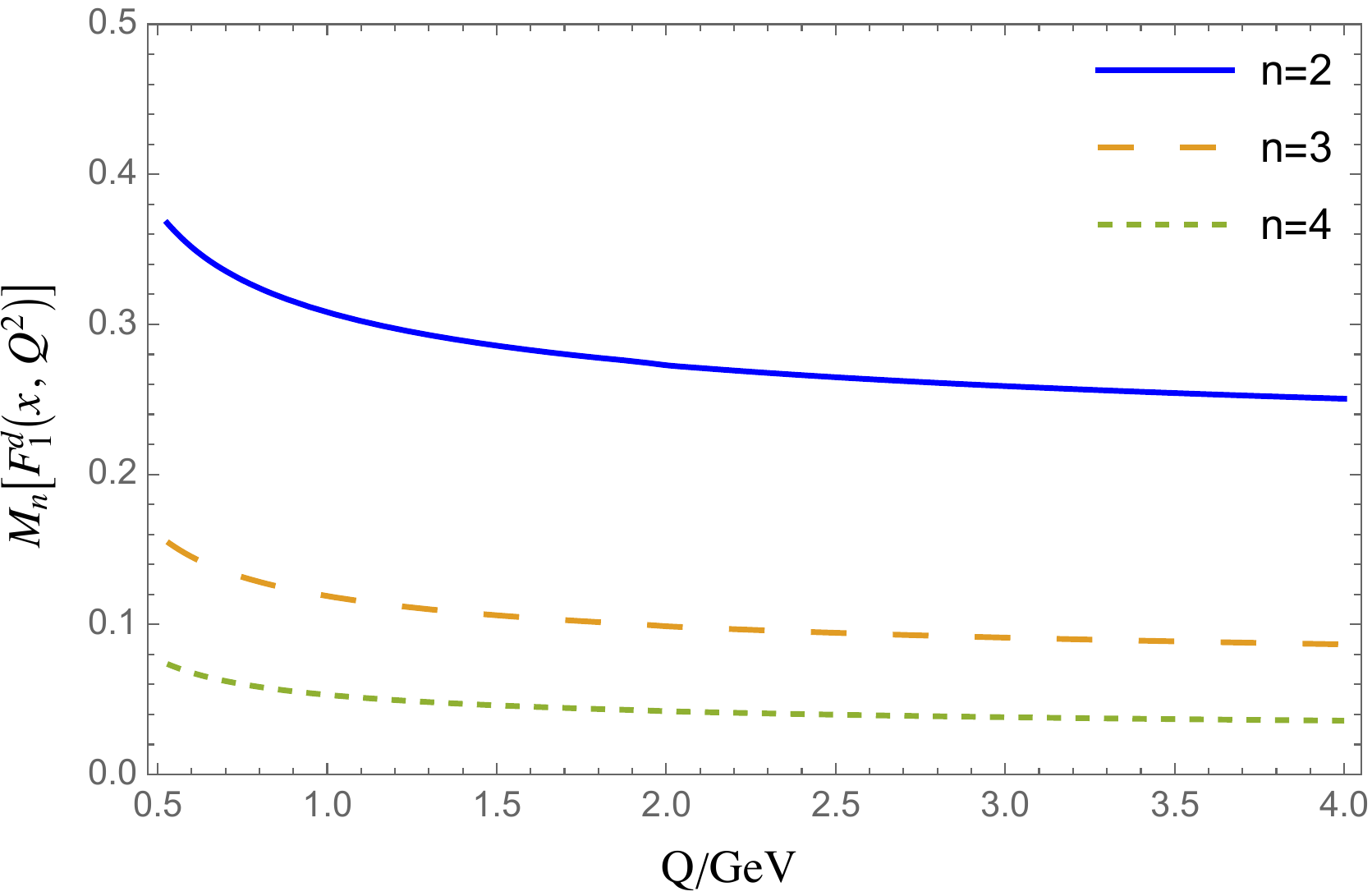}
	\caption{Evolution of the Mellin moments of the d quark unpolarized structure function $F_1^d(x)$. }
	\label{figureEvolution}
\end{figure}

As shown in Fig.~\ref{figureEvolution}, the contributions of the struck $d$ quark in the moments of the unpolarized structure function $F_1^d(x)$ decrease as the scale increases.  
The first moment ($n=2$) of $F_1^d(x)$ indicates the longitudinal momentum fraction carried by the $d$ quark which is around $37\%$. The rest of the momentum of the $\Delta$ is carried by the diquark (which is two $u$ quarks in this case) at the quark model scale. After the evolution from the quark-model scale to the scale $Q=4$ GeV, the percentage $37\%$ decreases to around 25\%. Correspondingly the sum of the fractions carried by the quark and the diquark is approximately 75\% at the scale $Q=4$ GeV. The rest fraction is distributed to gluons and sea quarks including $s$ and $c$ quarks and antiquarks.
As far as we know there is no experimental data or lattice QCD calculation for the moments of the structure functions.   
Thus, our prediction remains to be checked. 

\section{Summary and discussions}\label{summary}

Following the definitions of  GPDs for the spin-3/2 particle, this work has done the numerical calculation using $\Delta^+$ as an example for the first time. The results of other $\Delta$ isobars, $\Delta^-$, $\Delta^0$ and $\Delta^{++}$, can be obtained similarly.
Employing the diquark spectator approach, the unpolarized and polarized GPDs have been given.
A comparison of the electromagnetic and gravitational intrinsic properties with the previous work shows that this diquark spectator approximation is still reasonable.

The numerical GPDs are then plotted as functions of the light-cone loop momentum fraction $x$ and the minus transfer momentum square $-t$ as 3D figures.
We know that the GPDs are defined flavor by flavor, and thus, they are explained by utilizing the $d$ quark part of $\Delta^+$ as a typical example.
We see that the GPDs satisfy the time reversal relations. To illustrate the structures of the GPDs in particular, the 2D sectional planes at the associated $-t$ are given. It is found that the maximums or minimums shift to the large $x$ as $\lvert \xi \rvert$ increases and the maximums or minimums of the GPDs $H_{1,2,5,6}$ and $\tilde{H}_{5,6}$, which are related to form factors, are close to the position, $x = \frac{m_q+m_D \lvert \xi \rvert}{m_q+m_D}$. In addition, the axial vector form factors have also been given in comparison to the lattice QCD.
Moreover, our axial charge is $g_A=0.727$ which is consistent with other models or the lattice QCD in general.
In the forward limit, the structure functions $F_1$, $b_1$, $g_1$ and $g_2$ can be obtained from GPDs, and $F_1$ and $g_1$ have the maximum at the position, $x_{\text{max}} \sim \frac{m_q+m_D \lvert \xi \rvert}{m_q+m_D}$. The higher scale distributions of these structure functions are predicted after the evolution with the help of the QCDNUM package. We expect those results would be a reference for future experiments.

\section*{Acknowledgments}
The authors thank Jambul Gegelia for useful discussions. This work is supported by the National Natural 
Science Foundation of China under Grants No. 11975245, 
No. 11947224, No. 11947228, and No. 12035007. This work is 
also supported by the Sino-German CRC 110 “Symmetries and the 
Emergence of Structure in QCD” project by NSFC under Grant
No. 12070131001, the Key Research Program of Frontier Sciences, 
CAS, under Grant No. Y7292610K1, and the National Key Research 
and Development Program of China under Contracts 
No. 2020YFA0406300, and Guangdong Provincial funding with 
Grant No. 2019QN01X172, 
Guangdong Major Project of
Basic and Applied Basic Research No. 2020B0301030008, and the Department of Science and Technology of Guangdong Province with Grant No. 2022A0505030010. This work is supported in part by the MKW NRW under the funding code NW21-024-A and by the EU Horizon 2020 research and innovation programme (STRONG-2020, grant agreement No. 824093).

\appendix
\renewcommand\thesection{Appendix~\Alph{section}}
\section{Some useful identities}\label{appendixuseful}
\vspace{0.2cm} \par\noindent\par\setcounter{equation}{0}
\renewcommand{\theequation}{A\arabic{equation}}

In this section, many identities will be derived from Refs.~\cite{Cotogno:2019vjb,Fu:2022rkn,Fu:2022bpf}.
Before deriving the useful on-shell identities, one major general identity is the Schouten identity:
\begin{equation}\label{schoutenidentity}
  i \epsilon^{\mu \nu \rho \tau} g^{\lambda \beta} + i \epsilon^{\lambda \mu \nu \rho} g^{\tau \beta} + i \epsilon^{\tau \lambda \mu \nu} g^{\rho \beta} +
  i \epsilon^{\rho \tau \lambda \mu} g^{\nu \beta} + i \epsilon^{\nu \rho \tau \lambda} g^{\mu \beta} = 0.
\end{equation}
For any antisymmetric tensor $A^{\mu \nu}$, there is
\begin{equation}\label{schoutenidentityanti}
  2 i \epsilon_\sigma^{\,\, \mu \nu \rho} A^{\sigma \lambda} = i
  \epsilon^{\tau \sigma \mu \nu} A_{\tau \sigma} g^{\rho \lambda} + i
  \epsilon^{\tau \sigma \rho \mu} A_{\tau \sigma} g^{\nu \lambda} + i
  \epsilon^{\tau \sigma \nu \rho} A_{\tau \sigma} g^{\mu \lambda} .
\end{equation}
And 
\begin{equation}\label{appendixcepsigma}
  i \sigma^{\mu \nu} \gamma^5 = - \frac{i}{2} \epsilon^{\rho \sigma \mu \nu} i \sigma_{\rho \sigma},
\end{equation}
Eq.~\eqref{schoutenidentityanti} leads then to
\begin{equation} \label{B22a}
  - i \epsilon_\sigma^{\,\, \mu \nu \rho} i
  \sigma^{\sigma \lambda} = i \sigma^{\mu \nu} \gamma^5 g^{\rho
  \lambda} + i \sigma^{\rho \mu} \gamma^5 g^{\nu \lambda} + i \sigma^{\nu
  \rho} \gamma^5 g^{\mu \lambda}.
\end{equation}
According to the convention of the momenta, there are some on-shell conditions~\cite{Cotogno:2019vjb},
\begin{subequations}
  \begin{align}
    P^\alpha \doteq \frac{q^\alpha}{2}, & \quad
    P^{\alpha '} \doteq - \frac{q^{\alpha '}}{2}, \label{appendixcmomenta}\\
    \gamma^\alpha \doteq 0, & \quad \gamma^{\alpha '} \doteq 0, \label{appendixcgamma}
  \end{align}
\end{subequations}
where $\doteq$ represents the on-shell equality like the well-known Gordon identity.
From the Dirac equation, several on-shell relations were derived, and the Gordon identity,
\begin{equation}\label{gordonidentity}
  \gamma^{\mu} \doteq \frac{P^{\mu}}{M} + \frac{i \sigma^{\mu q}}{2 M},
\end{equation}
is the most famous one.
And some practical on-shell relations have been derived in Ref.~\cite{Cotogno:2019vjb},
\begin{subequations}
  \begin{align}
    \mathbbm{1} \doteq \frac{\slashed{P}}{M}, & \quad 0 \doteq \slashed{q},  \label{gordon1}\\
    \gamma^5 \doteq \frac{\slashed{q} \gamma^5}{2 M}, & \quad 0 \doteq \slashed{P} \gamma^5,\label{gordon5} \\
    \gamma^\mu \doteq \frac{P^\mu}{M} + \frac{i \sigma^{\mu q}}{2 M}, & \quad 0 \doteq \frac{q^\mu}{2} + i \sigma^{\mu P}, \label{gordon} \\
    \gamma^\mu \gamma^5 \doteq \frac{q^\mu \gamma^5}{ 2 M} + \frac{i \sigma^{\mu P} \gamma^5}{M}, & \quad 0 \doteq P^\mu \gamma^5 + \frac{i \sigma^{\mu q} \gamma^5}{2}, \label{appendixcb15} \\
    i \sigma^{\mu \nu} \gamma^5 \doteq - \frac{P^{[ \mu \nobracket} \gamma^{\nobracket \nu ]} \gamma^5}{M}+\frac{i \epsilon^{\mu \nu q \lambda} \gamma_\lambda}{2 M} , & \quad 0 \doteq - \frac{q^{[ \mu \nobracket} \gamma^{\nobracket \nu ]} \gamma^5}{2} + i \epsilon^{\mu \nu P \lambda} \gamma_\lambda, \label{appendixcb17} \\
    i \epsilon^{\rho \mu \alpha \lambda} \gamma_\lambda \doteq g^{\mu \alpha} \gamma^\rho \gamma^5 - g^{\rho \alpha} \gamma^\mu \gamma^5 ,& \quad i \epsilon^{\alpha' \mu \sigma \lambda} \gamma_\lambda \doteq g^{\alpha' \mu} \gamma^\sigma \gamma^5 - g^{\alpha' \sigma} \gamma^\mu \gamma^5. \label{appendixcb27}
  \end{align}
\end{subequations}
Contracting the Schouten identity \eqref{schoutenidentity} with $P_\rho q_\tau \gamma_\beta$, one can obtain using Eq.~\eqref{gordon1}
\begin{equation}
  - i \epsilon^{P q \mu \nu} \gamma^{\lambda} + M i \epsilon^{\lambda q \mu \nu}
  \doteq i \epsilon^{\lambda P q [\mu \nobracket} \gamma^{\nobracket \nu]} .
  \label{schouten2}
\end{equation}
Contracting this relation with $g^{\alpha}_{\mu} g^{\alpha'}_{\nu} n_{\lambda}$ gives
then using Eq.~\eqref{gordonidentity}
\begin{equation}
  -i \epsilon^{P q \alpha \alpha'}
  i \sigma^{n q} \doteq 2 i \epsilon^{P q \alpha \alpha'} P \cdot n - 2 M^2 i \epsilon^{n q \alpha \alpha'} .
\end{equation}
Contracting Eq.~\eqref{schouten2} with $g^{\alpha}_{\nu} n_{\mu} n_{\lambda}$
gives
\begin{equation}
  - i \epsilon^{P q n \alpha} \slashed{n} + M i \epsilon^{n q n \alpha} \doteq
  i \epsilon^{n P q n} \gamma^{\alpha} - i \epsilon^{n P q \alpha} \slashed{n}.
\end{equation}
The following two useful relations have been derived in Ref.~\cite{Fu:2022rkn},
\begin{equation} \label{sigma21}
  - P^{\alpha'} g^{\mu \alpha} - P^{\alpha} g^{\mu \alpha'} \doteq  - \frac{1}{M} \left(2 P^{\alpha'} P^{\alpha} - P^2 g^{\alpha' \alpha}\right) \gamma^{\mu} - g^{\alpha' \alpha} P^{\mu},
\end{equation}
and
\begin{equation} \label{sigma22}
  \begin{split}
    &- 2 P^{\alpha'} g^{\alpha \mu} i \sigma^{\nu q} - 2 P^{\alpha}
    g^{\alpha' \mu} i \sigma^{\nu q} \\
    \doteq & q^2 g^{\mu \nu} g^{\alpha' \alpha} + 8 g^{\mu \nu} P^{\alpha'}
    P^{\alpha} - g^{\alpha' \alpha} P^{\mu} i \sigma^{\nu q} + 2 g^{\alpha'
    \alpha} P^{\mu} P^{\nu} - 2 P^{\{ \alpha' \nobracket} g^{\nobracket \alpha
    \} \{ \mu \nobracket} P^{\nobracket \nu \}}
    - g^{\alpha' \alpha} q^{\mu} q^{\nu} - 2 P^{[\alpha' \nobracket}
    g^{\nobracket \alpha] \{ \mu \nobracket} q^{\nobracket \nu \}} \\
    &- q^2 g^{\alpha' \{ \mu \nobracket} g^{\nobracket \nu \} \alpha} - \frac{q^2}{2
    P^2} P^{\{ \alpha' \nobracket} g^{\nobracket \alpha \} [\mu \nobracket}
    P^{\nobracket \nu]}
    - \frac{2}{P^2} P^{\alpha'} P^{\alpha} P^{\mu} i \sigma^{\nu q} -
    \frac{4}{P^2} P^{\alpha'} P^{\alpha} P^{\mu} P^{\nu} + \frac{2 M}{P^2}
    P^{\nu}  (2 P^{\alpha'} P^{\alpha} - P^2 g^{\alpha' \alpha}) \gamma^{\mu}.
  \end{split} 
\end{equation}
Combining Eqs. \eqref{sigma21} and \eqref{sigma22}, one can eliminate the last
term in Eq. \eqref{sigma22},
\begin{equation}
  \begin{aligned}
    &- 2 P^{\alpha'} g^{\mu \alpha} i \sigma^{\nu q} - 2 P^{\alpha} g^{\mu
    \alpha'} i \sigma^{\nu q} \\
    \doteq & q^2 g^{\mu \nu} g^{\alpha' \alpha} + 8 g^{\mu \nu} P^{\alpha'}
    P^{\alpha} - g^{\alpha' \alpha} P^{\mu} i \sigma^{\nu q} + 2 g^{\alpha'
    \alpha} P^{\mu} P^{\nu} - 2 P^{\{ \alpha' \nobracket} g^{\nobracket \alpha
    \} \{ \mu \nobracket} P^{\nobracket \nu \}}
    - g^{\alpha' \alpha} q^{\mu} q^{\nu} - 2 P^{[\alpha' \nobracket}
    g^{\nobracket \alpha] \{ \mu \nobracket} q^{\nobracket \nu \}} \\
    & - q^2 g^{\alpha' \{ \mu \nobracket} g^{\nobracket \nu \} \alpha} - \frac{q^2}{2
    P^2} P^{\{ \alpha' \nobracket} g^{\nobracket \alpha \} [\mu \nobracket}
    P^{\nobracket \nu]}
    - \frac{2}{P^2} P^{\alpha'} P^{\alpha} P^{\mu} i \sigma^{\nu q} -
    \frac{4}{P^2} P^{\alpha'} P^{\alpha} P^{\mu} P^{\nu}\\
    & + \frac{2 M^2}{P^2} P^{\nu} (P^{\alpha'} g^{\mu \alpha} + P^{\alpha}
    g^{\mu \alpha'}) - \frac{2 M^2}{P^2} g^{\alpha' \alpha} P^{\mu}.
  \end{aligned}
\end{equation}
Contracting this on-shell identity with $ i \epsilon^{P q n}_{\quad \,\,\,\,\mu} n_{\nu}$ gives
\begin{equation}\label{appendixcc12}
  - 2 i \epsilon^{P q n \{ \alpha \nobracket} P^{\nobracket \alpha' \}} i
  \sigma^{n q} \doteq - 2 \left(q \cdot n\right) i \epsilon^{P q n [\alpha \nobracket} P^{\alpha' \nobracket]}  - q^2 i \epsilon^{P q n \{ \alpha \nobracket} n^{\nobracket \alpha' \}}.
\end{equation}
Moreover, contracting the Schouten identity \eqref{schoutenidentity} with $P_\mu q_\nu n_\rho P_\beta g^\alpha_{\,\, \tau} g^{\alpha'}_{\,\, \lambda}$, one can derive
\begin{equation}
  i \epsilon^{P q n [\alpha \nobracket} P^{\nobracket
  \alpha']} = - \left(P \cdot n\right) i \epsilon^{P q \alpha \alpha'}  + i \epsilon^{n q \alpha \alpha'} P^2.
\end{equation}
And whereupon Eq.~\eqref{appendixcc12} becomes
\begin{equation}
  i \epsilon^{P q n \{ \alpha
  \nobracket} P^{\nobracket \alpha' \}} i \sigma^{n q} \doteq - \left(P \cdot n\right) \left(q \cdot n\right) i \epsilon^{P q \alpha \alpha'} + P^2 \left(q \cdot n\right) i \epsilon^{n q
  \alpha \alpha'} + \frac{1}{2} q^2 i \epsilon^{P q n \{ \alpha \nobracket}
  n^{\nobracket \alpha' \}}.
\end{equation}

Changing indix $\rho$ in the first one of Eq.~\eqref{appendixcb27} to $\alpha'$ and using
Eq.~\eqref{appendixcgamma}, one can get
\begin{equation}\label{appendixc:16}
  i \epsilon^{\alpha' \mu \alpha \lambda} \gamma_{\lambda} \doteq - g^{\alpha' \alpha} \gamma^{\mu} \gamma^5 .
\end{equation}
One can therewith contract this relation with $n_{\mu}$, $P_{\mu}$ or $q_{\mu}$ and using Eq.~\eqref{gordon5}, there are
\begin{subequations}
  \begin{align}
    i \epsilon_{n \alpha' \alpha \lambda} \gamma^{\lambda} & \doteq 
    g_{\alpha' \alpha} \slashed{n} \gamma^5,\\
    i \epsilon_{\alpha' P \alpha \lambda} \gamma^{\lambda} & \doteq 0,\\
    i \epsilon_{\alpha' q \alpha \lambda} \gamma^{\lambda} & \doteq - 2 M
    g_{\alpha' \alpha} \gamma^5.
  \end{align} \label{B21a}
\end{subequations}
Contracting Eq.~\eqref{appendixcb27} with $n_{\rho} P_{\mu}  (n_{\sigma} P_{\mu})$, $n_{\rho}
q_{\mu}  (n_{\sigma} q_{\mu})$ or $P_{\rho} q_{\mu}  (P_{\sigma} q_{\mu})$ and using Eqs.~\eqref{gordon1}-\eqref{gordon5}, one get then
\begin{subequations}
  \begin{align}
    i \epsilon^{n P \alpha \lambda} \gamma_{\lambda} & \doteq P^{\alpha}
    \slashed{n} \gamma^5 , & \quad \epsilon^{\alpha' P n \lambda} \gamma_{\lambda} & \doteq P^{\alpha'}
    \slashed{n} \gamma^5, \\
    i \epsilon^{n q \alpha \lambda} \gamma_{\lambda} & \doteq q^{\alpha}
    \slashed{n} \gamma^5 - 2 M n^{\alpha} \gamma^5 , & \quad
    i \epsilon^{\alpha' q n \lambda} \gamma_{\lambda} & \doteq q^{\alpha'}
    \slashed{n} \gamma^5 - 2 M n^{\alpha'} \gamma^5, \label{appendixc18b} \\
    i \epsilon^{P q \alpha \lambda} \gamma_{\lambda} & \doteq - 2 M
    P^{\alpha} \gamma^5 , & \quad
    i \epsilon^{\alpha' q P \lambda} \gamma_{\lambda} & \doteq - 2 M
    P^{\alpha'} \gamma^5.
  \end{align} \label{B20a}
\end{subequations}
Combining Eqs.~\eqref{gordon5} and \eqref{appendixcb17} contracted with $q_\mu n_\nu$
, one can obtain
\begin{equation} \label{epsga1}
  i \epsilon^{P q n \lambda} \gamma_{\lambda} \doteq \frac{q^2 \slashed{n}
  \gamma^5}{2} - M \left(q \cdot n\right) \gamma^5 .
\end{equation}
Contracting Eq.~\eqref{appendixcb15} with $n_{\mu}$, there are two relations,
\begin{equation}
  \slashed{n} \gamma^5 \doteq \frac{q \cdot n }{2 M} \gamma^5 + \frac{i \sigma^{n P} \gamma^5}{M}, \quad 0 \doteq \left(P \cdot n\right) \gamma^5 + \frac{i \sigma^{n q} \gamma^5}{2}.
\end{equation}

Now we will derive some on-shell identities including the form $\sigma^{\mu \nu}$.
% $\sigma^{A B} = \sigma^{\mu \nu} A_{\mu} B_{\nu} = \sigma_{\mu \nu} A^{\mu}
% B^{\nu} = \sigma_{A B}$, so
Using Eqs.~\eqref{gordon1} and \eqref{gordon5} and contracting \eqref{appendixcb17} with $( P_\mu, q_\mu, n_\mu, g^\alpha_\mu ) \otimes ( P_\nu, q_\nu, n_\nu, g^{\alpha'}_\nu )$, we have
\begin{subequations}\label{sigma5}
  \begin{align}
    i \sigma^{P q} \gamma^5 & \doteq - 2 P^2 \gamma^5, &
    i \sigma^{n q} \gamma^5 & \doteq - 2 \left(P \cdot n\right) \gamma^5,\\
    i \sigma^{\alpha P}
    \gamma^5 & \doteq - \frac{1}{2} q^{\alpha} \gamma^5, &
    i \sigma^{\alpha'
    P} \gamma^5 & \doteq - \frac{1}{2} q^{\alpha'} \gamma^5,\\
    i \sigma^{\alpha q} \gamma^5 & \doteq - 2 P^{\alpha}
    \gamma^5, &
    i \sigma^{\alpha' q} \gamma^5 & \doteq - 2 P^{\alpha'}
    \gamma^5,\\
    i \sigma^{n \alpha}
    \gamma^5 & \doteq n^{\alpha} \gamma^5, &
    i \sigma^{n
    \alpha'} \gamma^5 & \doteq - n^{\alpha'} \gamma^5,\\
    i \sigma^{n P} \gamma^5 & \doteq M
    \slashed{n} \gamma^5 - \frac{1}{2} \left(q \cdot n\right) \gamma^5, &
    i \sigma^{\alpha' \alpha} \gamma^5 & \doteq g^{\alpha' \alpha}
    \gamma^5 .
  \end{align}
\end{subequations}
Owing to Eq.~\eqref{appendixcepsigma}, one can get these relations using Eq.~\eqref{sigma5},
\begin{subequations}
  \begin{align}
    i \epsilon^{P q \sigma \rho} i \sigma_{\sigma \rho} & \doteq 4 P^2 \gamma^5, &\quad
    i \epsilon^{n q \sigma \rho} i \sigma_{\sigma \rho} & \doteq 4 \left(P \cdot n\right) \gamma^5,\\
    i \epsilon^{P \alpha \sigma \rho} i \sigma_{\sigma \rho} & \doteq - q^{\alpha} \gamma^5 , & \quad
    i \epsilon^{P \alpha' \sigma \rho} i \sigma_{\sigma \rho} & \doteq - q^{\alpha'} \gamma^5,\\
    i \epsilon^{q \alpha \sigma \rho} i \sigma_{\sigma \rho} & \doteq - 4 P^{\alpha} \gamma^5 , & \quad
    i \epsilon^{q \alpha' \sigma \rho} i \sigma_{\sigma \rho} & \doteq - 4 P^{\alpha'} \gamma^5,\\
    i \epsilon^{P n \sigma \rho} i \sigma_{\sigma \rho} & \doteq 2 M \slashed{n} \gamma^5 - \left(q \cdot n\right) \gamma^5 , & \quad
    i \epsilon^{\alpha' \alpha \sigma \rho} i \sigma_{\sigma \rho} & \doteq - 2 g^{\alpha' \alpha} \gamma^5.
  \end{align}
\end{subequations}
From Eq. \eqref{B22a}, one have
\begin{equation}
  i\epsilon^{\mu \nu \rho \sigma} i\sigma_{\sigma A} = i \sigma^{\mu \nu}
  \gamma^5 A^{\rho} + i \sigma^{\rho \mu} \gamma^5 A^{\nu} + i \sigma^{\nu
  \rho} \gamma^5 A^{\mu} . \label{B23a}
\end{equation}
If contracting Eq. \eqref{B23a} with $P_{\mu} q_{\nu} n_{\rho}$ (or $P_{\mu}
q_{\nu} g_{\rho}^{\alpha}$, $n_{\mu} q_{\nu} g_{\rho}^{\alpha}$, $P_{\mu}
n_{\nu} g_{\rho}^{\alpha}$, $P_{\mu} q_{\nu} g_{\rho}^{\alpha'}$, $n_{\mu}
q_{\nu} g_{\rho}^{\alpha'}$, $P_{\mu} n_{\nu} g_{\rho}^{\alpha'}$, $P_{\mu} g_{\nu}^{\alpha'} g_{\rho}^{\alpha}$, $q_{\mu} g_{\alpha'}^{\nu}
g_{\rho}^{\alpha}$, $n_{\mu} g_{\alpha'}^{\nu} g_{\rho}^{\alpha}$), one obtains
\begin{subequations}
  \begin{align}
    i \epsilon^{P q n \sigma} i\sigma^{\sigma A} &= i \sigma^{P q} \gamma^5 A
    \cdot n + i \sigma^{n P} \gamma^5 A \cdot q + i \sigma^{q n} \gamma^5
    A \cdot P,\\
    i\epsilon^{P q \alpha \sigma} i \sigma^{\sigma A} &= i \sigma^{P q} \gamma^5
    A^{\alpha} + i \sigma^{\alpha P} \gamma^5 A \cdot q + i \sigma^{q
    \alpha} \gamma^5 A \cdot P,\\
    i \epsilon^{n q \alpha \sigma} i \sigma^{\sigma A} &= i \sigma^{n q} \gamma^5
    A^{\alpha} + i \sigma^{\alpha n} \gamma^5 A \cdot q + i \sigma^{q
    \alpha} \gamma^5 A \cdot n,\\
    i \epsilon^{P n \alpha \sigma} i \sigma^{\sigma A} &= i \sigma^{P n} \gamma^5
    A^{\alpha} + i \sigma^{\alpha P} \gamma^5 A \cdot n + i \sigma^{n
    \alpha} \gamma^5 A \cdot P,\\
    i \epsilon^{P q \alpha' \sigma} i \sigma^{\sigma A} &= i \sigma^{P q} \gamma^5
    A^{\alpha'} + i \sigma^{\alpha' P} \gamma^5 A \cdot q + i \sigma^{q
    \alpha'} \gamma^5 A \cdot P,\\
    i \epsilon^{n q \alpha' \sigma} i \sigma^{\sigma A} &= i \sigma^{n q} \gamma^5
    A^{\alpha'} + i \sigma^{\alpha' n} \gamma^5 A \cdot q + i \sigma^{q
    \alpha'} \gamma^5 A \cdot n,\\
    i \epsilon^{P n \alpha' \sigma} i \sigma^{\sigma A} &= i \sigma^{P n} \gamma^5
    A^{\alpha'} + i \sigma^{\alpha' P} \gamma^5 A \cdot n + i \sigma^{n
    \alpha'} \gamma^5 A \cdot P,\\
    i \epsilon^{P \alpha' \alpha \sigma} i \sigma^{\sigma A} &= i \sigma^{P
    \alpha'} \gamma^5 A^{\alpha} + i \sigma^{\alpha P} \gamma^5 A^{\alpha'} +
    i \sigma^{\alpha' \alpha} \gamma^5 A \cdot P,\\
    i \epsilon^{q \alpha' \alpha \sigma} i \sigma^{\sigma A} &= i \sigma^{q
    \alpha'} \gamma^5 A^{\alpha} + i \sigma^{\alpha q} \gamma^5 A^{\alpha'} +
    i \sigma^{\alpha' \alpha} \gamma^5 A \cdot q,\\
    i \epsilon^{n \alpha' \alpha \sigma} i \sigma^{\sigma A} &= i \sigma^{n
    \alpha'} \gamma^5 A^{\alpha} + i \sigma^{\alpha n} \gamma^5 A^{\alpha'} +
    i \sigma^{\alpha' \alpha} \gamma^5 A \cdot n,
  \end{align}
\end{subequations}
and $A$ can be changed to $n$ (or $P$, $q$).

From Eqs.~\eqref{B21a} and \eqref{B20a}, using the Schouten identity \eqref{schoutenidentity}, we can get
\begin{subequations}
  \begin{align}
    2 M g^{\alpha' \alpha} \gamma^5 & \doteq  - \frac{P_{\mu} P^{\mu}}{P^2} i
    \epsilon^{\alpha' q \alpha \lambda} \gamma_{\lambda} \doteq  \frac{M}{P^2} i \epsilon^{P \alpha' q
    \alpha} + \frac{1}{P^2} i \epsilon^{\lambda P \alpha' q} \gamma_{\lambda}
    P^{\alpha} + \frac{1}{P^2} i \epsilon^{q \alpha \lambda P}
    \gamma_{\lambda} P^{\alpha'}, \label{eqspq1}\\
    2 M g^{\alpha' \alpha} \gamma^5 & \doteq - \frac{n_{\mu} P^{\mu}}{P
    \cdot n} i \epsilon^{\alpha' q \alpha \lambda} \gamma_{\lambda}
    \doteq \frac{M}{P \cdot n} i \epsilon^{n
    \alpha' q \alpha} + \frac{1}{P \cdot n} i \epsilon^{\lambda n \alpha'
    q} \gamma_{\lambda} P^{\alpha} + \frac{1}{P \cdot n} i \epsilon^{q
    \alpha \lambda n} \gamma_{\lambda} P^{\alpha'} ,
  \end{align}
\end{subequations}
and
\begin{subequations}\label{eqspq2}
  \begin{align}
    2 M n^{\alpha} \gamma^5 - q^{\alpha} \slashed{n} \gamma^5
    & \doteq - \frac{P_{\mu} P^{\mu}}{P^2} i \epsilon^{n q \alpha \lambda}
    \gamma_{\lambda} \doteq \frac{M}{P^2} i \epsilon^{P n q \alpha} +
    \frac{1}{P^2} i \epsilon^{\lambda P n q} \gamma_{\lambda} P^{\alpha} +
    \frac{P \cdot n}{P^2} i \epsilon^{q \alpha \lambda P} \gamma_{\lambda} , \\
    q^{\alpha'} \slashed{n} \gamma^5 - 2 M n^{\alpha'} \gamma^5
    & \doteq - \frac{P_{\mu} P^{\mu}}{P^2} i \epsilon^{n q \alpha' \lambda}
    \gamma_{\lambda}
    \doteq \frac{M}{P^2} i \epsilon^{P n q \alpha'} +
    \frac{1}{P^2} i \epsilon^{\lambda P n q} \gamma_{\lambda} P^{\alpha'} +
    \frac{P \cdot n}{P^2} i \epsilon^{q \alpha' \lambda P} \gamma_{\lambda} .
  \end{align}
\end{subequations}
Using Eqs.~\eqref{B20a} and \eqref{epsga1}, one obtains
\begin{subequations}\label{appendixc28}
  \begin{align}
    i \epsilon^{P \alpha' q \alpha} & \doteq 2 P^2 g^{\alpha' \alpha} \gamma^5 - 4 P^{\alpha'} P^\alpha \gamma^5, \label{appendixc28a}\\
    i \epsilon^{n \alpha' q \alpha} & \doteq 2 \left(P \cdot n\right) g^{\alpha' \alpha} \gamma^5 - 2 n^{\{ \alpha' \nobracket} P^{\nobracket \alpha \}} \gamma^5, \\
    i \epsilon_{P n q \alpha} & \doteq 2 P^2 n_{\alpha} \gamma^5 -
    2 M P_{\alpha} \slashed{n} \gamma^5 + \left(q \cdot n\right) P_{\alpha} \gamma^5 - 2 \left(P \cdot n\right) P_{\alpha}
    \gamma^5, \label{eqspq3} \\
    i \epsilon_{P n q \alpha'} & \doteq - 2 P^2 n_{\alpha'} \gamma^5 -
    2 M P_{\alpha'}  \slashed{n} \gamma^5 + \left(q \cdot n\right) P_{\alpha'} \gamma^5 + 2 \left( P \cdot n \right) P_{\alpha'}
    \gamma^5. \label{eqspq4}
  \end{align}
\end{subequations}
Further, using Eqs.~\eqref{schoutenidentity} and \eqref{appendixc28}, we get another relation,
\begin{equation}
  i \epsilon^{P n \alpha \alpha'} 
  \doteq \frac{4 P^2}{t}  (n^{\alpha} P^{\alpha'} - n^{\alpha'}
  P^{\alpha}) \gamma^5 - \frac{8 M}{t} P^{\alpha} P^{\alpha'}
  \slashed{n} \gamma^5 + \frac{2 P^2 \left(q \cdot n\right)}{t} g^{\alpha' \alpha} \gamma^5 .
\end{equation}\\

Analogously, combining Eqs.~\eqref{B21a} and \eqref{B20a} and the Schouten identity \eqref{schoutenidentity}, one can obtain
\begin{subequations}
  \begin{align}
    q^{\alpha'} \slashed{n} \gamma^5 - 2 M n^{\alpha'} \gamma^5
    \doteq & - \frac{P_{\mu} n^{\mu}}{P \cdot n} i \epsilon^{n q \alpha' \lambda} \gamma_{\lambda}
    = \frac{1}{P \cdot n}  i \epsilon^{P n q \alpha'}
    \slashed{n} + \frac{1}{P \cdot n} i \epsilon^{\lambda P n q}
    \gamma_{\lambda} n^{\alpha'} + \frac{1}{P \cdot n} i \epsilon^{\alpha'
    \lambda P n} \gamma_{\lambda} q \cdot n,\\
    2 M n^{\alpha} \gamma^5 - q^{\alpha} \slashed{n} \gamma^5
    \doteq & - \frac{P_{\mu} n^{\mu}}{P \cdot n} i \epsilon^{n q \alpha
    \lambda} \gamma_{\lambda}
    = \frac{1}{P \cdot n}  i \epsilon^{P n q \alpha}
    \slashed{n} + \frac{1}{P \cdot n} i \epsilon^{\lambda P n q}
    \gamma_{\lambda} n^{\alpha} + \frac{1}{P \cdot n} i \epsilon^{\alpha
    \lambda P n} \gamma_{\lambda} q \cdot n,\\
    - g^{\alpha' \alpha} \slashed{n} \gamma^5
    \doteq & - \frac{q_{\mu} n^{\mu}}{q \cdot n} i \epsilon^{n \alpha'
    \alpha \lambda} \gamma_{\lambda}
    = \frac{1}{q \cdot n}  i \epsilon^{q n \alpha'
    \alpha} \slashed{n} + \frac{1}{q \cdot n} i \epsilon^{\lambda q n
    \alpha'} \gamma_{\lambda} n^{\alpha} + \frac{1}{q \cdot n} i
    \epsilon^{\alpha \lambda q n} \gamma_{\lambda} n^{\alpha'},\\
    - g^{\alpha' \alpha} \slashed{n} \gamma^5
    \doteq & - \frac{P_{\mu} n^{\mu}}{P \cdot n} i \epsilon^{n \alpha'
    \alpha \lambda} \gamma_{\lambda}
    = \frac{1}{P \cdot n}  i \epsilon^{P n \alpha'
    \alpha} \slashed{n} + \frac{1}{P \cdot n} i \epsilon^{\lambda P n
    \alpha'} \gamma_{\lambda} n^{\alpha} + \frac{1}{P \cdot n} i
    \epsilon^{\alpha \lambda P n} \gamma_{\lambda} n^{\alpha'},\\
    - 2 M g^{\alpha' \alpha} \gamma^5
    \doteq & - \frac{P_{\mu} n^{\mu}}{P \cdot n} i \epsilon^{q \alpha'
    \alpha \lambda} \gamma_{\lambda}
    = \frac{1}{P \cdot n}  i \epsilon^{P q \alpha' \alpha} \slashed{n} + \frac{n^\alpha}{P \cdot n} i \epsilon^{\lambda P q \alpha'} \gamma_{\lambda} + \frac{n^{\alpha'}}{P \cdot n} i \epsilon^{\alpha \lambda P q} \gamma_{\lambda} + \frac{q \cdot n}{P \cdot n} i \epsilon^{\alpha' \alpha \lambda P} \gamma_\lambda.
  \end{align}
\end{subequations}
Eqs.~\eqref{B20a} and \eqref{epsga1} give
\begin{subequations}
  \begin{align}
    i \epsilon^{P q n \alpha'} \slashed{n} & \doteq \left(2 P \cdot n + q \cdot n \right) P^{\alpha'} \slashed{n} \gamma^5 + M \left(2 P \cdot n - q \cdot n \right) n^{\alpha'} \gamma^5 + \frac{q^2}{2} n^{\alpha'} \slashed{n} \gamma^5, \\
    i \epsilon^{P q n \alpha} \slashed{n} & \doteq - M\left( 2 P \cdot n + q \cdot n \right) n^{\alpha} \gamma^5 +\left(2 P \cdot n - q \cdot n\right) P^{\alpha} \slashed{n} \gamma^5 + \frac{q^2}{2} n^{\alpha} \slashed{n} \gamma^5, \\
    i \epsilon^{n q \alpha'
    \alpha} \slashed{n} & \doteq \left(q \cdot n \right) g^{\alpha' \alpha} \slashed{n} \gamma^5 + 4 M n^{\alpha'} n^\alpha \gamma^5 + 2 P^{[ \alpha' \nobracket} n^{\nobracket \alpha ]} \slashed{n} \gamma^5, \\
    i \epsilon^{P n \alpha'
    \alpha} \slashed{n} & \doteq - \left(P \cdot n\right) g^{\alpha' \alpha} \slashed{n} \gamma^5 + P^{\{ \alpha' \nobracket} n^{\nobracket \alpha \}} \slashed{n} \gamma^5,\\
    i \epsilon^{P q \alpha' \alpha} \slashed{n} & \cdot - 2 M \left( P \cdot n \right) g^{\alpha' \alpha} \gamma^5 +2 M P^{\{ \alpha' \nobracket} n^{\nobracket \alpha \}} \gamma^5.
  \end{align}
\end{subequations}

And now we will derive two useful on-shell identities.
Because of Eq.~\eqref{appendixcgamma},
the structures $i \sigma^{\mu \nu} \gamma^5$ carrying an $\alpha$ or $\alpha'$ index can also be eliminated owing to
\begin{equation} \label{eqgamma3}
    i \sigma^{\alpha' \mu} \gamma^5 
    \doteq g^{\mu \alpha'} \gamma^5, \quad
    i \sigma^{\mu \alpha} \gamma^5 
    \doteq g^{\mu \alpha} \gamma^5 .
\end{equation}
Starting from the product of Dirac matrices,
\begin{equation}\label{appendixcproduct5}
  \begin{split}
    \gamma^{\rho} \gamma^{\mu} \gamma^{\nu} \gamma^{\sigma} \gamma^5 & =
    g^{\rho \mu} g^{\nu \sigma} \gamma^5 - g^{\rho \nu} g^{\mu \sigma}
    \gamma^5 + g^{\rho \sigma} g^{\mu \nu} \gamma^5 + i \epsilon^{\rho \mu \nu
    \sigma} \gamma^5 \gamma^5 - g^{\rho \mu} i \sigma^{\nu \sigma} \gamma^5 +
    g^{\rho \nu} i \sigma^{\mu \sigma} \gamma^5\\
    & - g^{\rho \sigma} i \sigma^{\mu \nu} \gamma^5 - g^{\nu \sigma} i
    \sigma^{\rho \mu} \gamma^5 + g^{\mu \sigma} i \sigma^{\rho \nu} \gamma^5 -
    g^{\mu \nu} i \sigma^{\rho \sigma} \gamma^5,
  \end{split}
\end{equation}
we change the index $\sigma$ to $\alpha$, and give then this relation using Eq.~\eqref{eqgamma3},
\begin{equation}
  0 \doteq \gamma^{\rho} \gamma^{\mu} \gamma^{\nu} \gamma^{\alpha} \gamma^5
  \doteq i \epsilon^{\rho \mu \nu \alpha} - g^{\rho \alpha} i \sigma^{\mu
  \nu} \gamma^5 - g^{\nu \alpha} i \sigma^{\rho \mu} \gamma^5 + g^{\mu \alpha} i \sigma^{\rho \nu} \gamma^5.
\end{equation}
So there is this identity,
\begin{equation} \label{eqalpha}
  i \epsilon^{\rho \mu \nu \alpha} \doteq g^{\rho \alpha} i \sigma^{\mu \nu}
  \gamma^5 + g^{\nu \alpha} i \sigma^{\rho \mu} \gamma^5 - g^{\mu \alpha} i
  \sigma^{\rho \nu} \gamma^5 .
\end{equation}
If we also change the index $\rho$ to $\alpha'$ and use Eq. \eqref{eqgamma3},
the Eq. \eqref{eqalpha} will become
\begin{equation} \label{eqaa1}
  i \epsilon^{\alpha' \mu \nu \alpha} \doteq g^{\alpha' \alpha} i \sigma^{\mu
  \nu} \gamma^5 + g^{\nu \alpha} g^{\alpha' \mu} \gamma^5 - g^{\mu \alpha}
  g^{\alpha' \nu} \gamma^5 .
\end{equation}
In the same way, there is another identity,
\begin{equation}\label{eqalpha1}
  i \epsilon^{\alpha' \mu \nu \sigma} \doteq g^{\alpha' \mu} i \sigma^{\nu
  \sigma} \gamma^5 - g^{\alpha' \nu} i \sigma^{\mu \sigma} \gamma^5 +
  g^{\alpha' \sigma} i \sigma^{\mu \nu} \gamma^5 ,
\end{equation}
by changing the index $\rho$ into $\alpha'$ in Eq.~\eqref{appendixcproduct5}.\\

Contracting another useful identity from the Schouten identity \eqref{schoutenidentity},
\begin{equation}
  0 = i \epsilon^{\alpha' \mu \alpha \lambda} P^{\nu} + i \epsilon^{\nu
  \alpha' \mu \alpha} P^{\lambda} + i \epsilon^{\lambda \nu \alpha' \mu}
  P^{\alpha} + i \epsilon^{\alpha \lambda \nu \alpha'} P^{\mu} + i
  \epsilon^{\mu \alpha \lambda \nu} P^{\alpha'},
\end{equation}
with $\gamma_{\lambda}$ and using Eqs. \eqref{appendixcb27} and \eqref{appendixc:16}, one will obtain
\begin{equation}\label{eqgamma6}
  0 \doteq - g^{\alpha' \alpha} \gamma^{\mu} \gamma^5 P^{\nu} + M i
  \epsilon^{\alpha' \mu \nu \alpha} + (g^{\alpha' \nu} \gamma^{\mu} \gamma^5
  - g^{\alpha' \mu} \gamma^{\nu} \gamma^5) P^{\alpha} + g^{\alpha' \alpha}
  \gamma^{\nu} \gamma^5 P^{\mu}
  + (g^{\nu \alpha} \gamma^{\mu} \gamma^5 - g^{\mu \alpha} \gamma^{\nu} \gamma^5) P^{\alpha'}.
\end{equation}
One can combine Eqs.~\eqref{eqaa1} and \eqref{eqgamma6} and get then
\begin{equation}
  \begin{split}
    & M (g^{\alpha' \alpha} i \sigma^{\mu \nu} \gamma^5 + g^{\nu \alpha}
    g^{\alpha' \mu} \gamma^5 - g^{\mu \alpha} g^{\alpha' \nu} \gamma^5)\\
    \doteq & g^{\alpha' \alpha} \gamma^{\mu} \gamma^5 P^{\nu} - (g^{\alpha'
    \nu} \gamma^{\mu} \gamma^5 - g^{\alpha' \mu} \gamma^{\nu} \gamma^5)
    P^{\alpha} - g^{\alpha' \alpha} \gamma^{\nu} \gamma^5 P^{\mu} - (g^{\nu
    \alpha} \gamma^{\mu} \gamma^5 - g^{\mu \alpha} \gamma^{\nu} \gamma^5)
    P^{\alpha'}.
  \end{split} \label{eqbb2}
\end{equation}
Contracting Eq.~\eqref{eqbb2} with $P_{\mu}$ and using Eqs.~\eqref{gordon5} and \eqref{appendixcb15} give
\begin{equation} \label{eqbb3}
  \begin{split}
    M \left( \frac{1}{2} g^{\alpha' \alpha} q^{\nu} \gamma^5 - M
    g^{\alpha' \alpha} \gamma^{\nu} \gamma^5 + g^{\nu \alpha} P^{\alpha'}
    \gamma^5 - P^{\alpha} g^{\alpha' \nu} \gamma^5 \right)
    \doteq 2 P^{\alpha'} P^{\alpha} \gamma^{\nu} \gamma^5 - P^2 g^{\alpha'
    \alpha} \gamma^{\nu} \gamma^5.
  \end{split}
\end{equation}
And we can contract Eq. \eqref{eqbb3} with $n_{\nu}$ and get an important relation,
\begin{equation}
  {n^{[\alpha' \nobracket} P^{\nobracket \alpha]} \gamma^5
  \doteq \frac{q \cdot n}{2} g^{\alpha' \alpha} \gamma^5 - \frac{2}{M} P^{\alpha'} P^{\alpha} \slashed{n} \gamma^5 - \frac{q^2}{4 M} g^{\alpha' \alpha}  \slashed{n} \gamma^5}.
\end{equation}
Using the Schouten identity \eqref{schoutenidentity}, we can write
\begin{equation}\label{appendixcschoutenproduct}
  - g^{[\alpha' \nobracket \{ \mu \nobracket} i \epsilon^{\nobracket \alpha]
  \nobracket \nu \} P q} = 2 g^{\mu \nu} i \epsilon^{P q \alpha' \alpha} -
  P^{\{ \mu \nobracket} i \epsilon^{\nobracket \nu \} q \alpha' \alpha} +
  q^{\{ \mu \nobracket} i \epsilon^{\nobracket \nu \} P \alpha' \alpha} .
\end{equation}
Contracting Eq.~\eqref{eqaa1} with $P_\mu q_\nu$, $q_\mu$ and $P_\mu$ and employing Eqs.~\eqref{gordon} and \eqref{appendixcb15}, one can get these relations,
\begin{subequations}
  \begin{align}
    i \epsilon^{P q \alpha' \alpha} & \doteq - 2 g^{\alpha' \alpha} P^2 \gamma^5 + P^{[\alpha' \nobracket}
    q^{\nobracket \alpha]} \gamma^5,\\
    i \epsilon^{\nu q \alpha' \alpha} & \doteq - 2 g^{\alpha' \alpha} P^{\nu} \gamma^5 + g^{\nu [\alpha'
    \nobracket} q^{\nobracket \alpha]} \gamma^5,\\
    i \epsilon^{\nu P \alpha' \alpha} & \doteq M g^{\alpha' \alpha} \gamma^{\nu} \gamma^5 - \frac{1}{2}
    g^{\alpha' \alpha} q^{\nu} \gamma^5 + g^{\nu [\alpha' \nobracket}
    P^{\nobracket \alpha]} \gamma^5 .
  \end{align}
\end{subequations}
Therefore, Eq.~\eqref{appendixcschoutenproduct} can be divided into
\begin{equation} \label{gamma51}
  \begin{split}
    - g^{[\alpha' \nobracket \{ \mu \nobracket} i \epsilon^{\nobracket \alpha]
    \nobracket \nu \} P q} & \doteq - 4 g^{\mu \nu} g^{\alpha' \alpha} P^2
    \gamma^5 + 2 g^{\mu \nu} P^{[\alpha' \nobracket} q^{\nobracket \alpha]}
    \gamma^5 + 4 g^{\alpha' \alpha} P^{\mu} P^{\nu} \gamma^5 - P^{\{ \mu
    \nobracket} g^{\nobracket \nu \} [\alpha' \nobracket} q^{\nobracket
    \alpha]} \gamma^5\\
    & + M g^{\alpha' \alpha} q^{\{ \mu \nobracket} \gamma^{\nobracket \nu
    \}} \gamma^5 - g^{\alpha' \alpha} q^{\mu} q^{\nu} \gamma^5 + q^{\{ \mu
    \nobracket} g^{\nobracket \nu \} [\alpha' \nobracket} P^{\nobracket
    \alpha]} \gamma^5 .
  \end{split}
\end{equation}
We can also simplify $- g^{[\alpha' \nobracket \{ \mu \nobracket} i \epsilon^{\nobracket \alpha]
\nobracket \nu \} P q}$ by another path, and
\begin{equation} \label{appcschoutenproduct2}
  - g^{[\alpha' \nobracket \{ \mu \nobracket} i \epsilon^{\nobracket \alpha ]
  \nobracket \nu \} P q} = - g^{ \alpha' \mu} i \epsilon^{ \alpha  \nu P q} + g^{\alpha \mu} i
  \epsilon^{ \alpha' \nu  P q} -
  g^{ \alpha' \nu} i \epsilon^{\alpha \mu P q} + g^{\alpha \nu} i \epsilon^{\alpha' \mu P q}.
\end{equation}
Contracting Eq.~\eqref{eqalpha} with $P_\rho q_\mu$ and Eq.~\eqref{eqalpha1} with $P_\sigma q_\mu$, we get
\begin{subequations}\label{appcsch2}
  \begin{align}
    i \epsilon^{\alpha \nu P q} & \doteq 2 P^2 g^{\nu \alpha} \gamma^5 - M q^{\alpha} \gamma^{\nu}
    \gamma^5 + \frac{1}{2} q^{\alpha} q^{\nu} \gamma^5 - 2 P^{\alpha} P^{\nu}
    \gamma^5,\\
    i \epsilon^{\alpha' \nu P q} &\doteq - 2 P^2 g^{\alpha' \nu} \gamma^5 + 2 P^{\alpha'} P^{\nu}
    \gamma^5 + M q^{\alpha'} \gamma^{\nu} \gamma^5 - \frac{1}{2} q^{\alpha'}
    q^{\nu} \gamma^5.
  \end{align}
\end{subequations}
According to Eqs.~\eqref{appcschoutenproduct2} and \eqref{appcsch2}, one can obtain
\begin{equation}\label{gamma52}
  - g^{[\alpha' \nobracket \{ \mu \nobracket} i \epsilon^{\nobracket
  \alpha] \nobracket \nu \} P q}
  \doteq - 4 P^2 g^{ \alpha' \{ \mu \nobracket}
  g^{\nobracket \nu \} \alpha} \gamma^5 - 2 M P^{[\alpha' \nobracket}
  g^{\nobracket \alpha] \{ \mu \nobracket} \gamma^{\nobracket \nu \}}
  \gamma^5 + P^{[\alpha' \nobracket} g^{\nobracket \alpha] \{ \mu
  \nobracket} q^{\nobracket \nu \}} \gamma^5 + 2 P^{\{ \alpha' \nobracket}
  g^{\nobracket \alpha \} \{ \mu \nobracket} P^{\nobracket \nu \}} \gamma^5 .
\end{equation}
Combining Eqs.~\eqref{gamma51} and \eqref{gamma52} and contracting them with
$n_{\mu} n_{\nu}$, we can get the another important relation,
\begin{equation}
  \begin{split}
    4 M n^{[\alpha' \nobracket} P^{\nobracket \alpha]} \slashed{n} \gamma^5
    \doteq & [4 (P \cdot n)^2 - (q \cdot n)^2] g^{\alpha' \alpha} \gamma^5
    + 2 M q \cdot n g^{\alpha' \alpha}  \slashed{n} \gamma^5
    + 4 q \cdot n n^{[\alpha' \nobracket} P^{\nobracket \alpha]}
    \gamma^5 \\
    & + 8 P^2 n^{\alpha'} n^{\alpha} \gamma^5 - 8 P \cdot n P^{\{ \alpha' \nobracket} n^{\nobracket \alpha \}} \gamma^5.
  \end{split}
\end{equation}

\renewcommand\thesection{Appendix~\Alph{section}}
\section{Some transformations for the integrals with $k$ }\label{appendixtransformation}
\vspace{0.2cm} \par\noindent\par\setcounter{equation}{0}
\renewcommand{\theequation}{B\arabic{equation}}

To integrate over $k$, scalar forms of $k$ are needed. However, the tensor forms, such as $k^\mu$, $k^\mu k^\nu$ and so on, are difficult to integrate directly. Therefore, we need to transform the vector integrals of $k$ to the scalar ones. Here take one $k$ as an example, the transformation is
\begin{equation}
	\int \frac{\text{d}^4 k \, f(k)}{\mathfrak{D}}k^\mu =  \frac{1}{ 4 \xi ^2 P^2+t } \left[\int \frac{\text{d}^4k \, f(k)}{\mathfrak{D}} A_{11} \frac{n^\mu}{P \cdot n} + \int \frac{\text{d}^4k \, f(k)}{\mathfrak{D}} A_{12} P^\mu + \int \frac{\text{d}^4k \, f(k)}{\mathfrak{D}} A_{13} q^\mu\right],
\end{equation}
where $f(k)$ and $A_{ij}$ are the scalar products including $k$. To make the equality more concise, we will omit the integral symbol and other scalar product $\frac{f(k)}{\mathfrak{D}}$, and we use $\doteqdot$ to represent the simple form
\begin{equation}
	k^\mu \doteqdot \frac{1}{ 4 \xi ^2 P^2+t } \left[A_{11} \frac{n^\mu}{P \cdot n} + A_{12} P^\mu + A_{13} q^\mu\right].
\end{equation}
Contract $n^\mu$, $P^\mu$ and $q^\mu$ respectively with the left and the right-hand sides, and then
\begin{equation}
	\left(
	\begin{matrix}
		k \cdot n \\ k \cdot P \\ k \cdot q
	\end{matrix}
	\right)
	\doteqdot \frac{1}{4 \xi ^2 P^2+t}\left( 
	\begin{matrix}
		0 & P \cdot n & q \cdot n\\
		1 & P^2 & 0 \\
		\frac{q \cdot n}{P \cdot n} & 0 & q^2
	\end{matrix}
	\right)
	\left(\begin{matrix}
		A_{11} \\ A_{12} \\ A_{13}
	\end{matrix}\right).
\end{equation}
And solving this equation, one can get
\begin{subequations}
	\begin{align}
		A_{11} &= t \left(k \cdot  P\right)-P^2 \left[2 \xi  \left(k \cdot  q\right)+t x\right],\\
		A_{12} &= 2 \xi  \left(k \cdot  q\right)+4 \xi ^2 \left(k \cdot  P\right)+t x,\\
		A_{13} &= k \cdot  q+2 \xi  \left(k \cdot  P\right)-2 \xi  x P^2.
	\end{align}
\end{subequations}

To describe the higher order $k$ tensor more succinctly, the symmetry symbol $\mathcal{S}[A]$ that means the symmetry form of $A$, such as $\mathcal{S}[g^{\mu \nu}]=g^{\mu \nu}$ and $\mathcal{S}[P^\mu q^\nu]=P^\mu q^\nu + P^\nu q^\mu$, is employed. 
Therefore, the second order $k$ tensor can be transformed as
\begin{equation}
	\begin{split}
	k^\mu k^\nu \doteqdot &
	\frac{1}{\left( 4 \xi ^2 P^2+t \right)^2}\mathcal{S}\left[A_{21} g^{\mu \nu}+A_{22} \frac{n^{\mu } n^{\nu }}{\left(P \cdot n\right)^2}+A_{23} \frac{P^{\mu } n^{\nu }}{P \cdot n}+A_{24} \frac{q^{\mu } n^{\nu }}{P \cdot n}+A_{25} P^{\mu } P^{\nu } + A_{26} q^{\mu } P^{\nu } +A_{27} q^{\mu } q^{\nu } \right],
	\end{split}
\end{equation}
where 
\begin{subequations}
	\begin{align}
		A_{21}=& 4\xi ^2 \left[ k^2 P^2- \left(k \cdot  P\right)^2\right]+ 4 \xi \left(k \cdot  q\right) \left(x P^2 - k \cdot  P\right)+\left[-\left(k \cdot  q\right)^2-2 t x \left(k \cdot  P\right)+t k^2+t x^2 P^2\right], \\
		A_{22}=& 4 \xi ^2 P^2 \left[P^2 \left(k \cdot  q\right)^2+t k^2 P^2-t \left(k \cdot  P\right)^2\right] + 8 \xi t P^2 \left(k \cdot  q\right) \left( x P^2 - k \cdot  P\right) \notag \\ & + \left[-4 t^2 x P^2 \left(k \cdot  P\right)+t^2 \left(k^2 P^2+\left(k \cdot  P\right)^2\right)-t P^2 \left(k \cdot  q\right)^2+2 t^2 x^2 \left(P^2\right)^2\right], \\
		A_{23}=& -8 \xi ^3 P^2 \left(k \cdot  q\right) \left(k \cdot  P\right) -4 \xi ^2 \left[P^2 \left(k \cdot  q\right)^2+t x P^2 \left(k \cdot  P\right)+t k^2 P^2-2 t \left(k \cdot  P\right)^2\right]  \notag \\
		& + 2 \xi t \left(k \cdot  q\right) \left(3  k \cdot  P- 4 x P^2 \right) + t \left[3 t x \left(k \cdot  P\right)-t k^2+ \left(k \cdot  q\right)^2-2 t x^2 P^2\right], \\
		A_{24}=& 8 \xi ^3 P^2 \left[ k^2 P^2- \left(k \cdot  P\right)^2\right] + 12 \xi ^2 P^2 \left(k \cdot  q\right) \left( x P^2 - k \cdot  P\right) \notag \\
		& + 2 \xi  \left[-2 P^2 \left(k \cdot  q\right)^2-4 t x P^2 \left(k \cdot  P\right)+t k^2 P^2+t \left(k \cdot  P\right)^2+2 t x^2 \left(P^2\right)^2\right] \notag \\
		& + t \left(k \cdot  q\right) \left( k \cdot  P- x P^2 \right), \\
		A_{25}=& 16 \xi ^4 \left[2 \left(k \cdot  P\right)^2- k^2 P^2\right] + 16 \xi ^3 \left(k \cdot  q\right) \left(2 k \cdot P - x P^2 \right) \notag \\
		& + 4 \xi ^2 \left[2 \left(k \cdot  q\right)^2+4 t x \left(k \cdot  P\right) - t k^2 - t x^2 P^2\right]+4 \xi  t x \left(k \cdot  q\right)+t^2 x^2, \\
		A_{26}=& 8 \xi ^3 \left[- x P^2 \left(k \cdot  P\right)-  k^2 P^2+ 2 \left(k \cdot  P\right)^2\right] + 4 \xi ^2 \left(k \cdot  q\right) \left(4  k \cdot  P- 3 x P^2 \right) \notag \\
		& + \xi \left[4 \left(k \cdot  q\right)^2+6 t x \left(k \cdot  P\right)-2 t k^2-4 t x^2 P^2\right]+t x \left(k \cdot  q\right), \\
		A_{27}=& 4 \xi ^2 \left[ -2 x P^2 \left(k \cdot  P\right)- k^2 P^2+2 \left(k \cdot  P\right)^2+ x^2 \left(P^2\right)^2 \right]+ 8 \xi \left(k \cdot  q\right) \left(  k \cdot  P- x P^2 \right) \notag \\
		& + \left[2 \left(k \cdot  q\right)^2+2 t x \left(k \cdot  P\right)-t k^2-t x^2 P^2\right].
	\end{align}
\end{subequations}
Analogously, for the third order $k$ tensor,
\begin{equation}
	\begin{split}
		k^\mu k^\nu k^\omega \doteqdot &
		\frac{1}{\left( 4 \xi ^2 P^2+t \right)^3}\mathcal{S} \left[ \left( 4 \xi ^2 P^2+t \right) \left(A_{31} \frac{ n^{\mu } \bar{g}^{\nu  \omega }}{P \cdot n}+A_{32} P^{\mu } \bar{g}^{\nu  \omega }  +A_{33} q^{\mu } \bar{g}^{\nu  \omega } \right)+A_{34} \frac{ n^{\mu } n^{\nu } n^{\omega }}{\left(P \cdot n\right)^3}\right.\\
		&+A_{35} \frac{n^{\mu } n^{\nu } P^{\omega }}{\left(P \cdot n\right)^2}+A_{36} \frac{n^{\mu } n^{\nu } q^{\omega }}{\left(P \cdot n\right)^2}+A_{37} \frac{n^{\mu } P^{\nu } P^{\omega }}{P \cdot n}+A_{38} \frac{n^{\mu } P^{\nu } q^{\omega }}{\left(P \cdot n\right)}+A_{39} \frac{n^{\mu } q^{\nu } q^{\omega }}{\left(P \cdot n\right)}\\
		&\left.+A_{310} P^{\mu } P^{\nu } P^{\omega } +A_{311} q^{\mu } P^{\nu } P^{\omega }
		+A_{312} q^{\mu } P^{\nu } q^{\omega } +A_{313} q^{\mu } q^{\nu } q^{\omega } \right],
	\end{split}
\end{equation}
where
\begin{subequations}
	\begin{align}
		A_{31}=& 8 \xi ^3 P^2 \left(k \cdot  q\right) \left[ \left(k \cdot  P\right)^2- k^2 P^2\right] \notag \\
		& + 4 \xi ^2 \left[2 P^2 \left(k \cdot  q\right)^2 \left(k \cdot  P\right)-  t x k^2 \left(P^2\right)^2+  t x P^2 \left(k \cdot  P\right)^2- t \left(k \cdot  P\right)^3+  t k^2 P^2 \left(k \cdot  P\right)-2 x \left(P^2\right)^2 \left(k \cdot  q\right)^2\right] \notag \\
		& + 2 \xi \left(k \cdot  q\right) \left[ P^2 \left(k \cdot  q\right)^2- t k^2 P^2 - 2 t \left(k \cdot  P\right)^2- 3 t x^2 \left(P^2\right)^2 + 6 t x P^2  \left(k \cdot  P\right)\right] \notag \\
		& + t^2 \left[3  x^2 P^2 \left(k \cdot  P\right) - x k^2 P^2-2 x \left(k \cdot  P\right)^2 + k^2 \left(k \cdot  P\right) - x^3 \left(P^2\right)^2\right] + t \left(k \cdot  q\right)^2 \left[- k \cdot  P + x P^2 \right] , \\
		A_{32}=& 16 \xi ^4 \left(k \cdot  P\right) \left[k^2 P^2 - \left(k \cdot  P\right)^2\right] + 8 \xi ^3 \left(k \cdot  q\right) \left[ k^2 P^2  - 3 \left(k \cdot  P\right)^2 + 2 x P^2  \left(k \cdot  P\right)\right] \notag \\
		& +4 \xi ^2 \left[-3 \left(k \cdot  q\right)^2 \left(k \cdot  P\right)+t x^2 P^2 \left(k \cdot  P\right)+t x k^2 P^2-3 t x \left(k \cdot  P\right)^2+t k^2 \left(k \cdot  P\right)+2 x P^2 \left(k \cdot  q\right)^2\right]\notag \\
		& + 2 \xi \left(k \cdot  q\right) \left[- \left(k \cdot  q\right)^2+3 t x^2 P^2 - 4 t x  \left(k \cdot  P\right)+ t k^2 \right] \notag \\
		& + t \left[-2 t x^2 \left(k \cdot  P\right)+t x k^2 - x \left(k \cdot  q\right)^2 + t x^3 P^2\right], \\
		A_{33}=& -8 \xi ^3 \left[ \left(k \cdot  P\right)^2 -k^2 P^2 \right] \left(k \cdot P -x P^2 \right) \notag \\
		& + 4 \xi ^2 \left(k \cdot  q\right) \left[ k^2 P^2 - \left(k \cdot  P\right)^2 - 2 \left(k \cdot  P -x P^2\right)^2 \right] \notag \\
		& + 2 \xi  \left[-2 t x \left(k \cdot  P\right)  + t x^2 P^2  + t k^2 - 3 \left(k \cdot  q\right)^2 \right] \left(k \cdot  P - x P^2\right) \notag \\
		& + \left(k \cdot  q\right) \left[-\left(k \cdot  q\right)^2+t x^2 P^2 -2 t x  \left(k \cdot  P\right)+t k^2 \right] ,\\
		A_{34}=& 8 \xi ^3 \left(P^2\right)^2 \left(k \cdot  q\right) \left[- P^2 \left(k \cdot  q\right)^2 - 3 t k^2 P^2 + 3 t \left(k \cdot  P\right)^2\right] \notag \\
		& + 12 \xi ^2 t P^2 \left[- t x k^2 \left(P^2\right)^2+ t x P^2 \left(k \cdot  P\right)^2- t \left(k \cdot  P\right)^3+ t k^2 P^2 \left(k \cdot  P\right) \right] \notag \\
		& +36 \xi ^2 t \left(P^2\right)^2 \left(k \cdot  q\right)^2 \left(  k \cdot  P- x P^2 \right) \notag \\
		& + 6 \xi t P^2 \left(k \cdot  q\right) \left[- t k^2 P^2 - 6 t  \left(k \cdot  P\right)^2- 4 t x^2 \left(P^2\right)^2 + 8 t x P^2  \left(k \cdot  P\right) + P^2 \left(k \cdot  q\right)^2\right] \notag \\
		& + 3 t^2 P^2 \left[ 3 t x^2 P^2 \left(k \cdot  P\right)- t x k^2 P^2 - 2 t x \left(k \cdot  P\right)^2+ t k^2 \left(k \cdot  P\right)- \left(k \cdot  q\right)^2 \left(k \cdot  P\right) + x P^2 \left(k \cdot  q\right)^2 - t^2 x^3 \left(P^2\right)^2\right] \notag \\
		& + t^3 \left(k \cdot  P -  x P^2 \right)^3 ,\\
		A_{35}= & 16 \xi ^4 P^2 \left(k \cdot  P\right) \left[ P^2 \left(k \cdot  q\right)^2 - t \left(k \cdot  P\right)^2 + t k^2 P^2 \right] \notag \\
		& + 8 \xi ^3 P^2 \left(k \cdot  q\right) \left[P^2 \left(k \cdot  q\right)^2-7 t \left(k \cdot  P\right)^2+ 3 t k^2 P^2 + 4 t x P^2 \left(k \cdot  P\right)\right] \notag \\
		& + 4 \xi ^2 t^2 \left[2 x^2 \left(P^2\right)^2 \left(k \cdot  P\right)+ 3 x k^2 \left(P^2\right)^2- 7 x P^2 \left(k \cdot  P\right)^2+ 3 \left(k \cdot  P\right)^3- k^2 P^2 \left(k \cdot  P\right) \right] \notag \\
		& + 36 \xi ^2 t P^2 \left(k \cdot  q\right)^2 \left(- k \cdot  P+ x P^2\right) \notag \\
		& + 2 \xi t \left(k \cdot  q\right) \left[3 t k^2 P^2 + 5 t \left(k \cdot  P\right)^2+ 12 t x^2 \left(P^2\right)^2 - 20 t x P^2  \left(k \cdot  P\right)- 3 P^2 \left(k \cdot  q\right)^2\right] \notag \\
		& + t^2 \left[-10 t x^2 P^2 \left(k \cdot  P\right)+3 t x k^2 P^2+5 t x \left(k \cdot  P\right)^2-2 t k^2 \left(k \cdot  P\right)+2  \left(k \cdot  q\right)^2 \left(k \cdot  P\right)-3 x P^2 \left(k \cdot  q\right)^2+4 t x^3 \left(P^2\right)^2\right] ,\\
		A_{36}=& 32 \xi ^4 \left(P^2\right)^2 \left(k \cdot  q\right) \left[ \left(k \cdot  P\right)^2- k^2 P^2\right] \notag \\
		& + 8 \xi^3 P^2 \left[5 P^2 \left(k \cdot  q\right)^2  +3 t k^2 P^2 - 3 t \left(k \cdot  P\right)^2  \right] \left(k \cdot  P - x P^2 \right)  \notag \\
		& + 4 \xi ^2 P^2 \left(k \cdot  q\right) \left[3 P^2 \left(k \cdot  q\right)^2 + t\left( \left(k \cdot  P\right)^2 - k^2 P^2 \right) - 10 t \left(k \cdot  P - x P^2 \right)^2 \right] \notag \\
		& + 2 \xi  \left[ 4 t^2 x^2 \left(P^2\right)^2 + t^2 \left(k \cdot  P\right)^2 - 8 t^2 x P^2 \left(k \cdot  P\right)- 7 t P^2 \left(k \cdot  q\right)^2 \right] \left(k \cdot  P - x P^2\right) \notag \\
		& + \left(k \cdot  q\right) \left[ t^2 k^2 P^2 - t^2 \left(k \cdot  P\right)^2 + 2 t^2  \left(k \cdot  P - x P^2\right)^2-t P^2 \left(k \cdot  q\right)^2\right] ,\\
		A_{37}=& 32 \xi ^5 P^2 \left(k \cdot  q\right) \left[ k^2 P^2 - 2  \left(k \cdot  P\right)^2\right] \notag \\
		& + 16 \xi ^4 \left[ - 4 P^2 \left(k \cdot  q\right)^2 \left(k \cdot  P\right)+ t x k^2 \left(P^2\right)^2- 2 t x P^2 \left(k \cdot  P\right)^2 + 4 t \left(k \cdot  P\right)^3- 3 t k^2 P^2 \left(k \cdot  P\right) + 2 x \left(P^2\right)^2 \left(k \cdot  q\right)^2\right] \notag \\
		& + 8 \xi ^3 \left(k \cdot  q\right) \left[- 2 P^2 \left(k \cdot  q\right)^2-  t k^2 P^2 +10 t \left(k \cdot  P\right)^2+ 3 t x^2 \left(P^2\right)^2 - 14 t x P^2  \left(k \cdot  P\right)\right] \notag \\
		& + 4 \xi ^2 \left[- 7 t^2 x^2 P^2 \left(k \cdot  P\right)-  t^2 x k^2 P^2 + 10 t^2 x \left(k \cdot  P\right)^2-3 t^2 k^2 \left(k \cdot  P\right) + 8 t \left(k \cdot  q\right)^2 \left(k \cdot  P\right)-8 t x P^2 \left(k \cdot  q\right)^2 + t^2 x^3 \left(P^2\right)^2\right] \notag \\
		&  + 2 \xi t \left(k \cdot  q\right) \left[-9 t x^2 P^2 +10 t x \left(k \cdot  P\right)-2 t k^2 +2 \left(k \cdot  q\right)^2\right] \notag \\
		&  + 2 t^2 \left[ t x^2 \left(k \cdot  P\right) - t x k^2 + x \left(k \cdot  q\right)^2\right] + 3 t^3 x^2 \left( k \cdot  P -  x P^2\right) , \\
		A_{38}=& 32 \xi ^5 P^2 \left(k \cdot  P\right) \left[ k^2 P^2 - \left(k \cdot  P\right)^2\right] \notag \\
		& + 16 \xi ^4 P^2 \left(k \cdot  q\right) \left[-2 \left(k \cdot  P\right)^2+ 2 k^2 P^2 -3 \left(k \cdot  P\right)\left(k \cdot  P -x P^2 \right)\right] \notag \\
		& + 8 \xi ^3 \left[ \left(3 t \left(k \cdot  P\right)^2 - 2 t x P^2 \left(k \cdot  P\right) - t k^2 P^2 - 5 P^2 \left(k \cdot  q\right)^2 \right) \left(k \cdot  P - x P^2\right) + 2 t x P^2 \left(k^2 P^2 - \left(k \cdot  P\right)^2 \right) - 2 P^2 \left(k \cdot  q\right)^2 \left(k \cdot  P\right) \right] \notag \\
		& + 4 \xi ^2 \left(k \cdot  q\right) \left(- 3 P^2 \left(k \cdot  q\right)^2 + t k^2 P^2 - t \left(k \cdot  P\right)^2 + 2 t \left(4 \left(k \cdot  P\right) - 5 x P^2 \right) \left(k \cdot  P - x P^2\right)\right) \notag \\
		& + 2 \xi \left[ t^2 x \left( k^2 P^2-\left(k \cdot  P\right)^2 \right) + \left(6 t^2 x \left(k \cdot  P\right) - 4 t^2 x^2 P^2 -2 t^2 k^2 + 5 t \left(k \cdot  q\right)^2 \right) \left(k \cdot  P - x P^2 \right)-2 t x P^2 \left(k \cdot  q\right)^2\right] \notag \\
		& + \left(k \cdot  q\right) \left[3 t^2 x \left(k \cdot  P - x P^2\right)+ t^2 x^2 P^2 -t^2 k^2 +t \left(k \cdot  q\right)^2\right] ,\\
		A_{39}=& 32 \xi ^4 P^2 \left( k^2 P^2 -\left(k \cdot  P\right)^2 \right)\left( k \cdot  P - x P^2 \right) \notag \\
		& + 8 \xi ^3 P^2 \left(k \cdot  q\right) \left[3 \left(k^2 P^2 - \left(k \cdot  P\right)^2 \right) - 5 \left(k \cdot  P - x P^2\right)^2 \right] \notag \\
		& + 4 \xi ^2 \left[-6 t x P^2 \left(k \cdot  P\right) +2 t \left(k \cdot  P\right)^2 + t k^2 P^2 + 3 t x^2 \left(P^2\right)^2 -10 P^2 \left(k \cdot  q\right)^2 \right]  \left( k \cdot  P - x P^2 \right)   \notag \\
		& + 2 \xi \left(k \cdot  q\right) \left[-4 P^2 \left(k \cdot  q\right)^2 + 3 t k^2 P^2 - 3 t \left(k \cdot  P\right)^2 + 7 t \left(k \cdot  P- x P^2 \right)^2\right] \notag \\
		& + \left[2 t^2 x \left(k \cdot  P\right) +2 t \left(k \cdot  q\right)^2 - t^2 x^2 P^2 -t^2 k^2 \right] \left(k \cdot  P - x P^2 \right) , \\
		A_{310}=& 64 \xi ^6 \left(k \cdot  P\right) \left[4 \left(k \cdot  P\right)^2-3 k^2 P^2 \right] + 96 \xi ^5 \left(k \cdot  q\right) \left[- k^2 P^2  + 4 \left(k \cdot  P\right)^2 - 2 x P^2 \left(k \cdot  P\right)\right]\notag \\
		& -48 \xi ^4 \left[-4 \left(k \cdot  q\right)^2 \left(k \cdot  P\right)+t x^2 P^2 \left(k \cdot  P\right)+t x k^2 P^2-4 t x \left(k \cdot  P\right)^2+t k^2 \left(k \cdot P\right)+2 x P^2 \left(k \cdot  q\right)^2\right] \notag \\
		& + 8 \xi ^3 \left(k \cdot  q\right) \left[4 \left(k \cdot  q\right)^2- 9 t x^2 P^2 +18 t x \left(k \cdot  P\right)- 3 t k^2 \right] \notag \\
		& + 12 \xi ^2 \left[3 t^2 x^2 \left(k \cdot  P\right)-  t^2 x k^2+ 2 t x \left(k \cdot  q\right)^2 -  t^2 x^3 P^2\right]  \notag \\
		&+6 \xi  t^2 x^2 \left(k \cdot  q\right)+t^3 x^3 , \\
		A_{311}=& 32 \xi ^5 \left[2\left(k \cdot  P\right) \left( \left(k \cdot  P\right)^2 -  k^2 P^2 \right)  + \left(2 \left(k \cdot  P\right)^2 - k^2 P^2 \right) \left(k \cdot  P - x P^2\right) \right] \notag \\
		& +16 \xi ^4 \left(k \cdot  q\right) \left[-3 k^2 P^2 + 12  \left(k \cdot  P\right)^2 + 2 x^2 \left(P^2\right)^2- 10 x P^2 \left(k \cdot  P\right)\right] \notag \\
		& + 8 \xi ^3 \left[4 \left(k \cdot  q\right)^2 \left( 3 k \cdot  P - 2 x P^2\right) - 7 t x^2 P^2 \left(k \cdot  P\right) - t x k^2 P^2 + 10 t x \left(k \cdot  P\right)^2 - 3 t k^2 \left(k \cdot  P\right) + t x^3 \left(P^2\right)^2\right] \notag \\
		& + 4 \xi ^2 \left(k \cdot  q\right) \left[4 \left(k \cdot  q\right)^2 - 9 t x^2 P^2 + 14 t x \left(k \cdot  P\right)- 3 t k^2\right] \notag \\
		& + 2 \xi  \left[5 t^2 x^2 \left(k \cdot  P\right)- 2 t^2 x k^2+ 4 t x \left(k \cdot  q\right)^2- 3 t^2 x^3 P^2\right] \notag \\
		& +t^2 x^2 \left(k \cdot  q\right) ,\\
		A_{312}=& 16 \xi ^4 \left[ x^2 \left(P^2\right)^2 \left(k \cdot  P\right) + 2 x k^2 \left(P^2\right)^2 - 4 x P^2 \left(k \cdot P\right)^2 + 4 \left(k \cdot  P\right)^3- 3 k^2 P^2 \left(k \cdot  P\right)\right] \notag \\
		& + 8 \xi ^3 \left(k \cdot  q\right) \left[-3 k^2 P^2 + 12 \left(k \cdot  P\right)^2 + 5 x^2 \left(P^2\right)^2 - 14 x P^2 \left(k \cdot  P\right)\right] \notag \\
		& + 4 \xi ^2 \left[12 \left(k \cdot  q\right)^2 \left(k \cdot  P\right)- 9 t x^2 P^2 \left(k \cdot  P\right)+ t x k^2 P^2+ 8 t x \left(k \cdot  P\right)^2-3 t k^2 \left(k \cdot  P\right)-10 x P^2 \left(k \cdot  q\right)^2+ 3 t x^3 \left(P^2\right)^2\right] \notag \\
		& + 2 \xi \left(k \cdot  q\right) \left[4 \left(k \cdot  q\right)^2- 7 t x^2 P^2 +10 t x \left(k \cdot  P\right)-3 t k^2 \right]\notag \\
		& + \left[2 t^2 x^2 \left(k \cdot  P\right)-t^2 x k^2+2 t x \left(k \cdot  q\right)^2-t^2 x^3 P^2\right] , \\
		A_{313}=& 8 \xi ^3 \left[ \left(k \cdot  P - x P^2\right)^2 + 3 \left(k \cdot  P\right)^2  -3 k^2 P^2 \right]\left(k \cdot  P - x P^2\right)
		+ 12 \xi ^2 \left(k \cdot  q\right) \left[\left(k \cdot  P\right)^2 - k^2 P^2 + 3  \left(k \cdot  P - x P^2\right)^2\right] \notag \\
		& + 6 \xi  \left[4 \left(k \cdot  q\right)^2  + 2 t x \left(k \cdot  P\right) - t x^2 P^2  -  t k^2 \right] \left(k \cdot  P - x P^2\right) \notag \\
		& + \left(k \cdot  q\right) \left(4 \left(k \cdot  q\right)^2 +6 t x  \left(k \cdot  P - x P^2\right) + 3 t x^2 P^2 -3 t k^2 \right) .
	\end{align}
\end{subequations}

\bibliographystyle{unsrt}
\bibliography{refGPDsNum}

\begin{thebibliography}{10}

\bibitem{Muller:1994ses}
Dieter M\"uller, D.~Robaschik, B.~Geyer, F.~M. Dittes, and J.~Ho\v{r}ej\v{s}i.
\newblock {Wave functions, evolution equations and evolution kernels from light
  ray operators of QCD}.
\newblock {\em Fortsch. Phys.}, 42:101--141, 1994.

\bibitem{Polyakov:2002yz}
M.~V. Polyakov.
\newblock {Generalized parton distributions and strong forces inside nucleons
  and nuclei}.
\newblock {\em Phys. Lett. B}, 555:57--62, 2003.

\bibitem{Ji:1996ek}
Xiang-Dong Ji.
\newblock {Gauge-Invariant Decomposition of Nucleon Spin}.
\newblock {\em Phys. Rev. Lett.}, 78:610--613, 1997.

\bibitem{Ji:1996nm}
Xiang-Dong Ji.
\newblock {Deeply virtual Compton scattering}.
\newblock {\em Phys. Rev. D}, 55:7114--7125, 1997.

\bibitem{Radyushkin:1997ki}
A.~V. Radyushkin.
\newblock {Nonforward parton distributions}.
\newblock {\em Phys. Rev. D}, 56:5524--5557, 1997.

\bibitem{CLAS:2018ddh}
M.~Hattawy et~al.
\newblock {Exploring the Structure of the Bound Proton with Deeply Virtual
  Compton Scattering}.
\newblock {\em Phys. Rev. Lett.}, 123(3):032502, 2019.

\bibitem{CLAS:2021gwi}
V.~Burkert et~al.
\newblock {Beam charge asymmetries for deeply virtual Compton scattering off
  the proton}.
\newblock {\em Eur. Phys. J. A}, 57(6):186, 2021.

\bibitem{Schnell:2012zz}
Gunar Schnell.
\newblock {Recent results on generalized parton distributions from the COMPASS,
  HERMES, and Jefferson Lab's Hall A and CLAS Collaborations}.
\newblock {\em Nucl. Phys. B Proc. Suppl.}, 222-224:187--198, 2012.

\bibitem{AbdulKhalek:2021gbh}
R.~Abdul~Khalek et~al.
\newblock {Science Requirements and Detector Concepts for the Electron-Ion
  Collider}: {EIC Yellow Report}.
\newblock {\em Nucl. Phys. A}, 1026:122447, 2022.

\bibitem{Anderle:2021wcy}
Daniele~P. Anderle et~al.
\newblock {Electron-ion collider in China}.
\newblock {\em Front. Phys. (Beijing)}, 16(6):64701, 2021.

\bibitem{Fu:2022bpf}
Dongyan Fu, Bao-Dong Sun, and Yubing Dong.
\newblock {Generalized parton distributions in spin-3/2 particles}.
\newblock {\em Phys. Rev. D}, 106(11):116012, 2022.

\bibitem{ParticleDataGroup:2022pth}
R.~L. Workman et~al.
\newblock {Review of Particle Physics}.
\newblock {\em PTEP}, 2022:083C01, 2022.

\bibitem{Alexandrou:2008bn}
C.~Alexandrou, T.~Korzec, G.~Koutsou, Th. Leontiou, C.~Lorce, J.~W. Negele,
  V.~Pascalutsa, A.~Tsapalis, and M.~Vanderhaeghen.
\newblock {Delta-baryon electromagnetic form factors in lattice QCD}.
\newblock {\em Phys. Rev. D}, 79:014507, 2009.

\bibitem{Fu:2022rkn}
Dongyan Fu, Bao-Dong Sun, and Yubing Dong.
\newblock {Electromagnetic and gravitational form factors of
  \ensuremath{\Delta} resonance in a covariant quark-diquark approach}.
\newblock {\em Phys. Rev. D}, 105(9):096002, 2022.

\bibitem{Pefkou:2021fni}
Dimitra~A. Pefkou, Daniel~C. Hackett, and Phiala~E. Shanahan.
\newblock {Gluon gravitational structure of hadrons of different spin}.
\newblock {\em Phys. Rev. D}, 105(5):054509, 2022.

\bibitem{Botje:2010ay}
M.~Botje.
\newblock {QCDNUM: Fast QCD Evolution and Convolution}.
\newblock {\em Comput. Phys. Commun.}, 182:490--532, 2011.

\bibitem{Lappi:2023lmi}
Tuomas Lappi, Heikki M\"antysaari, Hannu Paukkunen, and Mirja Tevio.
\newblock {Evolution of structure functions in momentum space}.
\newblock 4 2023.

\bibitem{Jaffe:1988up}
R.~L. Jaffe and Aneesh Manohar.
\newblock {Deep Inelastic Scattering from Arbitrary Spin Targets}.
\newblock {\em Nucl. Phys. B}, 321:343, 1989.

\bibitem{Nozawa:1990gt}
S.~Nozawa and D.~B. Leinweber.
\newblock {Electromagnetic form-factors of spin 3/2 baryons}.
\newblock {\em Phys. Rev. D}, 42:3567--3571, 1990.

\bibitem{Alexandrou:2010tj}
Constantia Alexandrou, Eric~B. Gregory, Tomasz Korzec, Giannis Koutsou, John
  Negele, Toru Sato, and Antonios Tsapalis.
\newblock {Axial and pseudoscalar form-factors of the $\Delta^+$(1232)}.
\newblock {\em PoS}, LATTICE2010:141, 2010.

\bibitem{Alexandrou:2013opa}
C.~Alexandrou, E.~B. Gregory, T.~Korzec, G.~Koutsou, J.~W. Negele, T.~Sato, and
  A.~Tsapalis.
\newblock {Determination of the $\Delta(1232)$ axial and pseudoscalar form
  factors from lattice QCD}.
\newblock {\em Phys. Rev. D}, 87(11):114513, 2013.

\bibitem{Jun:2020lfx}
Yu-Son Jun, Jung-Min Suh, and Hyun-Chul Kim.
\newblock {Axial-vector form factors of the baryon decuplet with flavor SU(3)
  symmetry breaking}.
\newblock {\em Phys. Rev. D}, 102(5):054011, 2020.

\bibitem{Diehl:2003ny}
M.~Diehl.
\newblock {Generalized parton distributions}.
\newblock {\em Phys. Rept.}, 388:41--277, 2003.

\bibitem{Kim:2020lrs}
June-Young Kim and Bao-Dong Sun.
\newblock {Gravitational form factors of a baryon with spin-3/2}.
\newblock {\em Eur. Phys. J. C}, 81(1):85, 2021.

\bibitem{Belitsky:2005qn}
A.~V. Belitsky and A.~V. Radyushkin.
\newblock {Unraveling hadron structure with generalized parton distributions}.
\newblock {\em Phys. Rept.}, 418:1--387, 2005.

\bibitem{Dong:2009yp}
Yubing Dong, Amand Faessler, Thomas Gutsche, Sergey Kovalenko, and Valery~E.
  Lyubovitskij.
\newblock {X(3872) as a hadronic molecule and its decays to charmonium states
  and pions}.
\newblock {\em Phys. Rev. D}, 79:094013, 2009.

\bibitem{Scadron:1968zz}
Michael~D. Scadron.
\newblock {Covariant Propagators and Vertex Functions for Any Spin}.
\newblock {\em Phys. Rev.}, 165:1640--1647, 1968.

\bibitem{Pauli:1949zm}
W.~Pauli and F.~Villars.
\newblock {On the Invariant regularization in relativistic quantum theory}.
\newblock {\em Rev. Mod. Phys.}, 21:434--444, 1949.

\bibitem{Sun:2017gtz}
Bao-Dong Sun and Yu-Bing Dong.
\newblock {$\rho$ meson unpolarized generalized parton distributions with a
  light-front constituent quark model}.
\newblock {\em Phys. Rev. D}, 96(3):036019, 2017.

\bibitem{Blumlein:1997pi}
Johannes Blumlein, Bodo Geyer, and Dieter Robaschik.
\newblock {On the evolution kernels of twist-2 light ray operators for
  unpolarized and polarized deep inelastic scattering}.
\newblock {\em Phys. Lett. B}, 406:161--170, 1997.

\bibitem{Golec-Biernat:1998zbo}
Krzysztof~J. Golec-Biernat and Alan~D. Martin.
\newblock {Off diagonal parton distributions and their evolution}.
\newblock {\em Phys. Rev. D}, 59:014029, 1999.

\bibitem{Kivel:1999wa}
N.~Kivel and L.~Mankiewicz.
\newblock {Conformal string operators and evolution of skewed parton
  distributions}.
\newblock {\em Nucl. Phys. B}, 557:271--295, 1999.

\bibitem{Sun:2018ldr}
Bao-Dong Sun and Yu-Bing Dong.
\newblock {Polarized generalized parton distributions and structure functions
  of the $\rho$ meson}.
\newblock {\em Phys. Rev. D}, 99(1):016023, 2019.

\bibitem{Cotogno:2019vjb}
Sabrina Cotogno, C\'edric Lorc\'e, Peter Lowdon, and Manuel Morales.
\newblock {Covariant multipole expansion of local currents for massive states
  of any spin}.
\newblock {\em Phys. Rev. D}, 101(5):056016, 2020.

\end{thebibliography}

\end{document}